\documentclass[sigconf, nonacm]{acmart}

\AtBeginDocument{%
  }

\usepackage{graphicx}
\usepackage[dvipsnames, table]{xcolor}
\usepackage[newfloat]{minted}
\usepackage{listings}
\usepackage{xspace}
\usepackage{textcomp}
\usepackage[]{hyperref}
\usepackage{xurl}
\usepackage{pifont}
\usepackage[ruled,linesnumbered]{algorithm2e}
\usepackage{amsmath}
\usepackage{amsfonts}
\usepackage{booktabs}
\usepackage{multirow}
\usepackage{multicol}
\usepackage{subfigure}
\usepackage{balance}
 \usepackage[labelfont=bf]{caption}
\usepackage{caption}
\usepackage{soul}
\usepackage[normalem]{ulem}
\usepackage{array}
\usepackage{threeparttable}
\usepackage{rotating}
\usepackage[most]{tcolorbox}
\usepackage{tikz}
\usetikzlibrary{tikzmark}
\usetikzlibrary{backgrounds}
\usepackage{enumitem}%
\usepackage{bbding}
\usepackage{wasysym}
\usepackage{lipsum}
\usepackage{breakurl}
\usepackage{url}

\hypersetup{
    colorlinks = true,
    linkcolor = Red,
    anchorcolor = Red,
    citecolor = Red,
    filecolor = Red,
    urlcolor = Red
}

\SetKwComment{Comment}{/* }{ */}
\SetKw{Continue}{continue}
\SetKw{Break}{break}
 
\graphicspath{{fig/}}

\definecolor{bg}{rgb}{0.95,0.95,0.95}
\colorlet{bgcolor}{yellow!12}
\colorlet{highlightcolor}{purple!16}
\definecolor{yetanothergreen}{RGB}{112, 173, 71}
\definecolor{yetanotherred}{RGB}{192, 0, 0}
\definecolor{maroon}{cmyk}{0,0.87,0.68,0.32}

\newcommand{\nop}[1]{}

\newcommand{\tabincell}[2]{\begin{tabular}{@{}#1@{}}#2\end{tabular}}
\newenvironment{packeditemize}{
\begin{list}{$\bullet$}{
\setlength{\itemsep}{1.5pt}
\setlength{\labelwidth}{8pt}
\setlength{\leftmargin}{10pt}
\setlength{\labelsep}{3pt}
\setlength{\listparindent}{\parindent}
\setlength{\parsep}{1.5pt}
\setlength{\parskip}{1.5pt}
\setlength{\topsep}{1.5pt}}
}{\end{list}}

\newtcolorbox[%
auto counter]{mybox}[2][]{%
	enhanced jigsaw,
        colback=yellow!12,
	breakable,
	#1}
\newtcolorbox[%
auto counter]{rqbox}[2][]{%
	enhanced jigsaw,
        colback=white!12,
	breakable,
	#1} 

\newcommand\featuretext[1]{
  \llap{\vrule width.35pt height2pt depth2.5pt\kern1pt}%
  \rlap{\rotatebox{40}{\textbf{#1}}}%
}
\newcommand{\cmark}{{\color{yetanothergreen}{\ding{51}}}}
\newcommand{\xmark}{{\color{yetanotherred}{\ding{55}}}}
\renewcommand{\paragraph}[1]{\vspace{0.05in}\noindent{\bf{#1}.}}

\newcommand{\circlenum}[1]{{\large{\ding{\numexpr171 + #1}}}}
\newcommand{\refappendix}[1]{\hyperref[#1]{Appendix~\ref*{#1}}}

\newcommand{\sys}{{\textsc{{FlexEmu}}}\xspace}

\begin{document}

\title{\sys: Towards Flexible MCU Peripheral Emulation \\(Extended Version)}

\author{Chongqing Lei}
\affiliation{
\institution{Southeast University}
\city{Nanjing}
\state{Jiangsu}
\country{China}
}
\email{leicq@seu.edu.cn}

\author{Zhen Ling}
\authornote{Corresponding Author}
\affiliation{
\institution{Southeast University}
\city{Nanjing}
\state{Jiangsu}
\country{China}
}
\email{zhenling@seu.edu.cn}

\author{Xiangyu Xu}
\affiliation{
\institution{Southeast University}
\city{Nanjing}
\state{Jiangsu}
\country{China}
}
\email{xy-xu@seu.edu.cn}

\author{Shaofeng Li}
\affiliation{
\institution{Southeast University}
\city{Nanjing}
\state{Jiangsu}
\country{China}
}
\email{shaofengli@seu.edu.cn}

\author{Guangchi Liu}
\affiliation{
\institution{Southeast University}
\city{Nanjing}
\state{Jiangsu}
\country{China}
}
\email{gc-liu@seu.edu.cn}

\author{Kai Dong}
\affiliation{
\institution{Southeast University}
\city{Nanjing}
\state{Jiangsu}
\country{China}
}
\email{dk@seu.edu.cn}

\author{Junzhou Luo}
\affiliation{
\institution{Southeast University}
\city{Nanjing}
\state{Jiangsu}
\country{China}
}
\email{jluo@seu.edu.cn}

\begin{abstract}
Microcontroller units (MCUs) are widely used in embedded devices due to their low power consumption and cost-effectiveness. MCU firmware controls these devices and is vital to the security of embedded systems. However, performing dynamic security analyses for MCU firmware has remained challenging due to the lack of usable execution environments -- existing dynamic analyses cannot run on physical devices (e.g., insufficient computational resources), while building emulators is costly due to the massive amount of heterogeneous hardware, especially peripherals. Recent advances in automated peripheral emulation have made MCU emulation more scalable. However, these efforts only support limited peripherals and are hard to extend because they require ad-hoc adaptations. \looseness=-1

Our work is based on the insight that MCU peripherals can be modeled in a two-fold manner. At the \textit{structural} level, peripherals have diverse implementations. But we can use a limited set of primitives to abstract peripherals because their hardware implementations are based on common hardware concepts. These primitives are abstract and can be instantiated with peripheral-specific implementation details to accommodate diverse peripheral implementations. At the \textit{semantic} level, peripherals have diverse functionalities. However, we can use a single unified semantic model to describe the same kind of peripherals because they exhibit similar functionalities. Primitives serve as basic building blocks, allowing flexible semantic model construction.
Building on this, we propose \sys, a flexible MCU peripheral emulation framework. Once semantic models are created, \sys automatically extracts peripheral-specific details to instantiate models and generate emulators accordingly.
We have successfully applied \sys to model 12 kinds of MCU peripherals. Our evaluation on 90 firmware samples across 15 different MCU platforms shows that the automatically generated emulators can faithfully replicate hardware behaviors and achieve a 98.48\% unit test passing rate, outperforming state-of-the-art approaches. To demonstrate the implications of \sys on firmware security, we use the generated emulators to fuzz three popular RTOSes and uncover 10 previously unknown bugs. \looseness=-1
\end{abstract}

\maketitle
\pagestyle{plain}
\thispagestyle{plain}

\section{Introduction}
\label{sec::intro}

Embedded devices have become ubiquitous, seamlessly integrated into every facet of daily life, from the Internet of Things (IoT) to industrial systems, revolutionizing the way we interact with the world.
At the heart of these embedded systems, microcontroller units (MCUs) are frequently adopted due to their compact design, low power consumption, and cost-effectiveness. The software that runs on MCUs, or MCU firmware, manages device operations and enables device interaction with the external environment.
Therefore, performing security analysis for MCU firmware becomes essential for the overall security of embedded devices \cite{Hernandez::FirmWire::NDSS22, Ruge::Frankenstein::Sec20, Komaromy::HuaweiTEE::BHUSA21, Ling::Plug::IOTJ17}.

Dynamic security analysis techniques have outstanding vulnerability discovery capabilities and thus have been applied to MCU firmware. However, this is particularly challenging due to the lack of usable execution environments.
Naturally, people have tried to directly perform dynamic analyses on physical devices \cite{Li::uAFL:ICSE22, Shi::IPEA::NDSS24, Mera::SHiFT::SEC24, Liu::CO3::SEC24}. However, this is infeasible for MCUs because existing dynamic analysis techniques (e.g., AddressSanitizer \cite{Serebryany::ASAN::ATC12}) typically consume massive computational resources and are designed for general-purpose platforms, while MCUs are resource-constrained and heterogeneous.
To overcome such constraints, emulation-based dynamic analysis techniques \cite{Bellard::QEMU::ATC05, Quynh::Unicorn::BHUSA15} become popular. However, emulated execution environments often do not exist for MCUs and are costly to build because MCUs use various instruction sets, and, more importantly, have diverse peripheral hardware implementations (e.g.,  >2,790 unique peripherals for ARM Cortex-based MCUs \cite{Fasano::SoK-Rehosting::AsiaCCS21}).

\textit{Rehosting} techniques have been proposed to facilitate emulation-based dynamic analysis of MCU firmware \cite{Zaddach:Avatar::NDSS14, Koscher::SURROGATES::WOOT15, Corteggiani::Inception::Sec18, Gustafson::Pretender::RAID19, Spensky::Conware::AsiaCCS21, Feng::P2IM::Sec20, Cao::Laelaps::ACSAC20, Mera::DICE::SP20, Zhou::uEmu::Sec21, Scharnowski::FuzzWare::Sec22, Zhou::sEmu::CCS22, Lei::Perry::Sec24}. It utilizes the computing resources of the host machine to create a controlled virtual execution environment (VXE) for MCU firmware, enabling advanced dynamic analysis techniques. However, hardware-based rehosting techniques orchestrate the VXE and the physical hardware through hardware-in-the-loop \cite{Zaddach:Avatar::NDSS14,Koscher::SURROGATES::WOOT15,Corteggiani::Inception::Sec18} or record-and-replay \cite{Gustafson::Pretender::RAID19, Spensky::Conware::AsiaCCS21} mechanisms and are not generally applicable in the absence of physical devices. As opposed to such methods, emulation-based rehosting techniques emulate the whole hardware platform to remove the dependency on physical devices. These methods are typically firmware dependent and suffer from limited universality and fidelity \cite{Feng::P2IM::Sec20, Mera::DICE::SP20, Cao::Laelaps::ACSAC20, Zhou::uEmu::Sec21, Scharnowski::FuzzWare::Sec22}. To further address this problem, firmware-agnostic emulation techniques have emerged in recent years \cite{Zhou::sEmu::CCS22, Lei::Perry::Sec24}. These techniques use heuristic rules to model a limited number of peripheral behaviors and extract these rules from hardware description materials to emulate peripherals.
However, such methods do not scale and only support limited peripherals -- their rules lack the expressiveness to describe all hardware behaviors and the extraction process requires material-specific adaptations.
Therefore, automatically building faithful emulators for diverse peripherals has remained challenging. \looseness=-1

Fortunately, we have gained an insight that peripherals can be modeled in a two-fold endeavor. First, peripheral hardware is built with common hardware concepts such as registers and interrupts, which implies that we can use a set of primitives to model a peripheral at the \textit{structural} level. These primitives are abstract and do not contain hardware implementation details, and can be instantiated with peripheral-specific implementation details. This design is compatible with diverse peripheral implementations. Peripherals also have diverse functionalities (or \textit{semantics}). However, peripherals of the same category (e.g., UART) share similar functionalities (e.g., transmit and receive data) and therefore can be modeled using a unified semantic model. Semantic models are composed of the above primitives, thus enabling flexible peripheral modeling.

Based on this insight, we propose a flexible MCU peripheral emulation framework, \sys. At its core, \sys provides a set of 9 primitives to build peripheral models. These primitives are designed based on our thorough investigation of MCU peripherals, covering basic hardware concepts like registers, interrupt events, and memory interactions. Semantic models are built using these abstract primitives, each with a short description to describe its hardware functionality.
Therefore, to generate peripheral-specific emulators, \sys consists of a frontend to extract peripheral-specific implementation details from peripheral description materials (i.e., driver source code) for model instantiation, and a template-based backend to generate emulators based on model instances. The generated emulators can be directly integrated into emulator frameworks such as QEMU \cite{Bellard::QEMU::ATC05}. \looseness=-1

The frontend of \sys faces a great challenge because extracting peripheral implementation details requires understanding high-level semantics of driver code. Luckily, the rise of large language models (LLMs) has made this process possible. \sys attaches a description to each primitive used by a model to describe their semantics (e.g., a data register), and asks the LLM to extract peripheral implementation details that are needed to instantiate the primitive from driver source code. However, the limitations of LLMs (e.g., hallucinations) pose a threat to the correctness of the extracted information. To address this challenge and achieve fully automated information extraction, we first adopt a code analysis-based scheme to correct ill-formed LLM responses. Then, we use domain knowledge to detect self-contradictions within a single response and between multiple responses and filter out invalid LLM responses. \looseness=-1

We implement a prototype of \sys with $\sim8,000$ lines of Python and C++ code. \sys is highly extensible and we have successfully applied it to model 12 kinds of commonly used MCU peripherals, most of which are not currently supported by QEMU.
We conduct a comprehensive evaluation of \sys on a dataset consisting of 90 firmware samples across 15 MCUs to assess whether the generated emulators can faithfully emulate various firmware. Evaluation results show that emulators generated by \sys can faithfully replicate hardware functionalities with a 98.48\% unit test passing rate in terms of execution fidelity, surpassing state-of-the-art approaches. We also illustrate the security implications of \sys. To this end, we use the generated emulators to fuzz the Bluetooth host stacks of three popular real-time operating systems (RTOS): Zephyr \cite{Zephyr::Zephyr::Web}, NuttX \cite{NuttX::NuttX::Web} and Mynewt/NimBLE \cite{Mynewt::Mynewt::Web}. As a result, we uncover a total of 10 previously unknown bugs.


In summary, we make the following major contributions:

\begin{packeditemize}
\item \textbf{Novel Insights.} We propose a two-fold peripheral modeling mechanism to facilitate automated peripheral emulation. The mechanism composes semantic models using primitives for the same kind of peripherals to accommodate diverse peripheral implementations and functionalities, thus greatly reducing the required emulation efforts.

\item \textbf{Practical Tool.} We introduce a flexible MCU peripheral emulation framework, \sys, to automatically generate emulators for MCU peripherals based on peripheral models, which uses a LLM-based frontend to extract model arguments for different MCU peripherals, and uses a backend to generate emulator code for them using template-based code synthesis.

\item \textbf{Extensive Evaluations.} We implement \sys to support 12 kinds of commonly used peripherals, use it to generate emulators for 15 MCUs and perform a thorough evaluation with 90 firmware samples to demonstrate the fidelity and universality of the generated emulators. \sys can generate faithful emulators to pass 98.48\% unit tests and emulate various firmware samples, outperforming the state-of-the-art approaches.

\item \textbf{Security Implications.} We use emulators generated by \sys to emulate and fuzz the Bluetooth host stacks of three popular RTOSes and uncover 10 previously unknown bugs.

\end{packeditemize}

\section{Background}
\label{sec::background}

\subsection{MCU-based Embedded Systems}
\label{subsec::background::mcu}
\begin{figure}[!htb]
    \centering
    \includegraphics[width=0.85\linewidth]{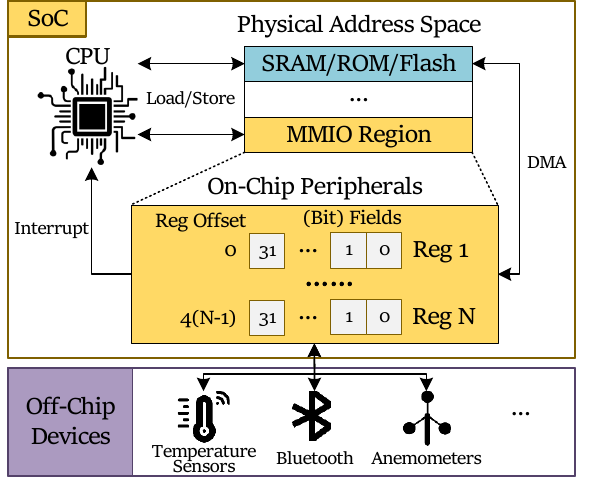}
    \small
    \caption{\textbf{Architecture of MCU-based embedded systems.}}
    \label{fig:mcu-arch}
\end{figure}

From a hardware perspective, MCU-based embedded systems typically consist of the MCU System-on-Chip (SoC) and various off-chip devices, as shown in \autoref{fig:mcu-arch}. The MCU SoC consists of a CPU, physical memory, and on-chip peripherals. The CPU runs a piece of software, i.e. the MCU firmware, to drive the embedded system. The on-chip peripherals, or peripherals\footnote{We use the two terms interchangeably in this paper.}, not only provide the only communication channels between the SoC and the outside world, but also offer various hardware-implemented functionalities to the CPU (e.g., crypto accelerators). These peripherals expose interfaces by mapping their registers into the MCU's physical address space, allowing the CPU to access registers through memory-mapped I/O (MMIO) when executing the firmware. Registers have various fields, which provide fine-grained access to hardware functionalities like configuring peripherals' working modes, acquiring their working status, and performing data exchanges, etc.
Peripherals are often connected to off-chip devices, such as various sensors, to help the system monitor and react to any environmental changes.

Besides being passively accessed by the CPU, peripherals can actively notify the CPU of the occurrence of certain events through interrupts. For example, when an Ethernet peripheral receives a packet, it triggers an interrupt to signal the CPU to process this packet. Peripherals may have multiple events, each with dedicated register fields to enable or disable the event and indicate its occurrence. These fields are used by the peripheral to generate interrupt signals. These signals are then transmitted via the peripheral’s outgoing interrupt lines to notify the CPU of the event. Due to the limited number of CPU interrupt ports, multiple events may share a single interrupt line. To manage interrupt lines, they are typically connected to an interrupt controller, which forwards interrupt signals to the CPU when the associated interrupt is raised and enabled.
Upon receiving an interrupt signal, the CPU pauses its current task and jumps to the corresponding interrupt service routine (ISR) to handle the interrupt.
To further enhance I/O performance, peripherals can bypass the CPU and exchange data directly with physical memory using direct memory access (DMA). This allows for faster data transfers and frees the CPU to handle other tasks. A dedicated kind of peripheral, i.e. DMA controllers, manages these data transfers independently and raises an interrupt to notify the CPU once a transmission completes.

\subsection{MCU-Oriented Rehosting Techniques}
\label{subsec::background::rehosting}
Efforts have been made to directly apply existing dynamic security analysis techniques to physical MCU-based devices \cite{Li::uAFL:ICSE22, Shi::IPEA::NDSS24, Mera::SHiFT::SEC24, Liu::CO3::SEC24}. However, this is infeasible because existing dynamic security analysis techniques consume substantial computational resources and are designed for general-purpose platforms, while MCUs have limited computing resources, as well as heterogeneous CPU architectures and peripherals.
Rehosting techniques have been proposed to address this challenge. These techniques use high-end machines to create virtual execution environments (VXEs) for MCU firmware. VXEs can leverage the host machine's abundant computing resources, thereby alleviating the resource limitations of the original hardware platform. To tackle the issue of heterogeneity, rehosting techniques must emulate both CPUs and peripherals \cite{Fasano::SoK-Rehosting::AsiaCCS21}, which typically demands significant manual efforts. Since the number of CPU models is relatively small, manually emulating them is generally acceptable. However, the huge number of peripherals has made manual emulation impractical -- a recent study by Fasano \textit{et al.} \cite{Fasano::SoK-Rehosting::AsiaCCS21} suggests that even only for ARM Cortex-based MCUs, there are more than 2,790 unique peripherals. Therefore, automated peripheral emulation becomes the key challenge for rehosting techniques. Existing rehosting techniques can be grouped into two categories based on their dependency on actual MCU hardware:

\noindent\textbf{Hardware-based Rehosting.} Some techniques leverage MCU hardware for MCU rehosting, and can be divided into two groups. The first kind of techniques is called hardware-in-the-loop \cite{Zaddach:Avatar::NDSS14, Koscher::SURROGATES::WOOT15, Corteggiani::Inception::Sec18}, where all peripheral accesses are forwarded (e.g., through debuggers) to the physical MCU. The other kind of techniques \cite{Gustafson::Pretender::RAID19, Spensky::Conware::AsiaCCS21} record peripheral interactions during on-device firmware execution, analyze peripheral responses, and replay them during emulated firmware execution. Hardware-based rehosting techniques still depend on physical hardware, thus facing scalability issues. \looseness=-1

\noindent\textbf{Emulation-based Rehosting.} Other techniques focus on building software-only VXEs to remove the dependency on physical hardware and can be divided into two groups:

\begin{packeditemize}
\begin{sloppypar}    
\item[\textbf{1) Firmware-Dependent Emulation:}] These techniques adapt the VXE for each firmware and can be achieved either at the driver level or register level.
Driver-level methods \cite{Clements::HALucinator::Sec20, Li::Para-rehosting::NDSS21, Seidel::SAFIREFUZZ::SEC23, Hofhammer::SURGEON::BAR24} manually write custom functions to mimic peripheral interaction semantics and replace the original driver functions within the target firmware. However, such approaches are less practical because they require adaptations for each driver library and even each firmware, which is both labor-intensive and error-prone.
Register-level methods \cite{Cao::Laelaps::ACSAC20, Feng::P2IM::Sec20, Scharnowski::FuzzWare::Sec22, Zhou::uEmu::Sec21} are specifically designed for firmware state exploration (e.g., fuzzing) instead of faithful emulation. To achieve this, they estimate proper values (e.g., via symbolic execution) for register accesses to bypass path constraints. Despite the limited applications, such approaches may also trigger spurious firmware states (e.g., false crashes) because they cannot replicate actual peripheral behaviors.
\end{sloppypar}
\item[\textbf{2) Firmware-Agnostic Emulation:}] These techniques faithfully emulate MCU hardware to emulate diverse MCU firmware and enable various analyses. Emulating hardware is typically achieved manually (e.g., QEMU \cite{Bellard::QEMU::ATC05}) and is labor-intensive. To automate this process, recent works use rules to model hardware behaviors and extract these rules from hardware description materials: SEmu \cite{Zhou::sEmu::CCS22} extracts hardware behaviors by analyzing hardware manuals, while Perry \cite{Lei::Perry::Sec24} does so by analyzing driver source code.
However, both solutions face extensibility challenges and support limited peripherals. First, their rules are limited in expressiveness and can describe only a narrow range of hardware behaviors. For example, SEmu rules cannot capture peripheral behaviors involving memory operations (e.g., manipulating Ethernet transfer descriptors). Besides, their implementations require material-specific adaptations (e.g., specific patterns in manuals) to function properly.
Therefore, automatically emulating a broader range of peripherals has remained a challenge.

\end{packeditemize}

\section{Problem Statement and Motivating Example}
\label{sec::overview}

\subsection{Problem Statement}
\label{subsec::problem-statement}
\noindent\textbf{Goals.} Our primary goal is to develop an extensible and flexible framework to automatically build faithful MCU peripheral emulators. The framework can be easily extended to support more peripherals and requires no adaptations to the used hardware description materials. As a benefit, we can automatically generate MCU emulators to facilitate dynamic firmware analysis.

However, achieving this goal is challenging. First, peripherals exhibit diverse functionalities. How can the framework be designed to model these functionalities more effectively than previous rule-based approaches? Second, even for peripherals with similar functionalities, their hardware implementations vary (e.g., STM32F4xx UART and FRDM-K64F UART). How can peripheral-specific implementation details be automatically extracted from hardware description materials to accommodate this diversity?



\noindent\textbf{Assumptions.} We make the following assumptions. First, like previous efforts \cite{Zhou::sEmu::CCS22, Lei::Perry::Sec24}, we assume that we have access to peripheral description materials, especially driver source code. MCU vendors typically provide high-quality driver source code that contains comprehensive information about peripheral hardware \cite{Lei::Perry::Sec24}. This source code can be utilized to generate faithful and firmware-agnostic hardware emulators \cite{Lei::Perry::Sec24}.
Second, we assume that we know the corresponding MCU of the target firmware so that we can obtain the corresponding driver code to extract peripheral implementation details. The MCU information can be obtained from product brochures, model numbers printed on the SoC, and strings or debugging information stored in the firmware \cite{Giese::XiaomiRev::DefCon26}. \looseness=-1

\subsection{Motivating Example}
\label{subsec::motivating_example}
Despite the challenges in automated peripheral emulation as discussed in \S\ref{subsec::problem-statement}, we have gained a key insight that allows us to achieve our goal.

\begin{mybox}[boxsep=0pt,
	boxrule=1pt,
	left=4pt,
	right=4pt,
 	top=4pt,
 	bottom=4pt,
	]
	
\textbf{Key Insight.} Scalable emulation of MCU peripherals can be achieved through a two-fold modeling mechanism. At the \textit{structural} level, peripherals can be modeled using a set of primitives because their hardware implementations are built with common hardware concepts like registers and interrupts. These primitives are abstract and can be instantiated with peripheral-specific information to accommodate diverse peripheral implementations. \textit{Semantic} level information is also required to replicate peripherals' functionalities. To this end, we find that peripherals that have similar functionalities can be described using a single unified semantic model. The semantic model consists of the above primitives, which serve as the model's parameters. Each used primitive is also attached with a description of its corresponding hardware functionality. Once the model is built, we can automatically emulate peripherals if peripheral-specific information can be extracted from driver source code to instantiate a concrete peripheral model. However, the extraction task is traditionally challenging because it requires understanding high-level code semantics. Luckily, the rise of large language models (LLMs) has made this task possible due to their powerful text and code understanding capabilities.
\end{mybox}

We illustrate our insight with a motivating example. Most MCUs have timer peripherals. For example, STM32F4xx MCUs have TIM and FRDM-K64F MCUs have FTM. Despite being implemented differently, these peripherals share common timer functionalities such as measuring elapsed ticks and raising periodic interrupts. To implement such functionalities, timers typically use registers to hold the current tick and period value. Once the current tick reaches the period value, an event is raised and specific register fields are set to indicate such an event. For example, STM32F4xx timers use the \texttt{CNT} register to hold the current tick value, the \texttt{ARR} register to hold the period value, and the \texttt{UIF} field of the \texttt{SR} register to indicate a timer event. Likewise, FRDM-K64F timers do so by using the \texttt{CNT} register, the \texttt{MOD} register, and the \texttt{TOF} field of the \texttt{SC} register. \looseness=-1

In the above example, there are two key primitives for timers -- register and register field. Based on the two primitives, we can use three parameters to construct a semantic model for timers: 
the tick register, the period register, and the event register field. With this model, the functionalities of timers can be emulated using a general code template as shown in \autoref{lst::timer-emu-template}.
Finally, we can automatically synthesize timer emulators by first obtaining peripheral-specific implementation details as concrete arguments (e.g., the actual register holding the tick value), then instantiate a concrete timer model (e.g., STM32F4xx timers) model the concrete arguments (e.g., \texttt{CNT}, \texttt{ARR} and \texttt{SR[UIF]}).

\begin{listing}[!htb]
\begin{minted}[
breakanywhere,
baselinestretch=1.0,
numbersep=2pt,
bgcolor=bgcolor,
fontsize=\scriptsize,
linenos,
breaklines,
highlightcolor=highlightcolor,
]{c}
// this gets executed periodically
timer->[tick] += 1;
if (timer->[tick] == timer->[period]) {
    SET_BIT(timer->[event field]);
}
\end{minted}
\small
\caption{\textbf{Pseudo code template for the timer emulator.}}
\label{lst::timer-emu-template}
\end{listing}

However, obtaining concrete model arguments is a challenging task because it requires the understanding of both parameter semantics and peripheral description materials. 
For example, how to denote that a register holds the tick value and how to decide the actual register based on the input materials? As discussed in \S\ref{subsec::background::rehosting}, prior solutions depend on ad-hoc knowledge about the used materials: SEmu \cite{Zhou::sEmu::CCS22} uses various keywords and phrases (e.g., \textit{``current timer value''}) that appeared in hardware manuals, while Perry \cite{Lei::Perry::Sec24} annotates specific driver functions or function parameters and performs program analysis. These solutions require users to be familiar not only with peripheral behaviors, but also with the used materials, thus significantly limiting their extensibility and practicality.

The emergence of LLMs has made text and code semantic reasoning practical, and can theoretically be applied to accomplish the above task without relying on additional knowledge about the used materials. Continuing with the above timer example, to decide the actual tick register, we can show the material (e.g., STM32F4xx driver source code) to the LLM and ask it to find \textit{``the register holding the current tick value''}, and the LLM may understand our instructions and return the desired register (i.e., \texttt{CNT}).

However, the limitations of current LLMs (e.g., hallucinations) make their responses unreliable. To address this issue, we can correct ill-formed responses with code analysis and filter out logically incorrect responses by detecting self-contradictions (e.g., returning a fake register).
Therefore, if we can orchestrate the above process, we can automatically emulate various peripherals.
This method is also more flexible and extensible as no knowledge about the used materials is required, and users only need to focus on peripherals.

\begin{figure*}[!htb]
    \centering
    \small
    \includegraphics[width=\linewidth]{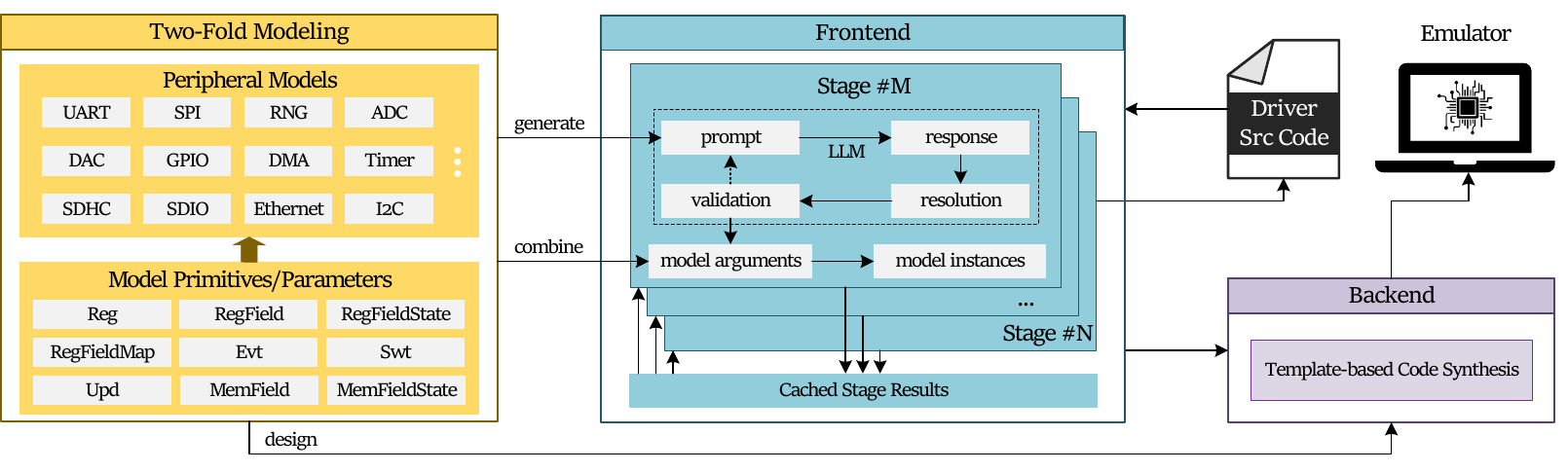}
    \caption{\textbf{\sys overview.}}
    \label{fig::design}
\end{figure*}

\section{Design}
\label{sec::design}

Following the insight and example in \S\ref{subsec::motivating_example}, we can conclude that to achieve our goal of flexible peripheral emulator construction, it is crucial to first properly build semantic models for each category of peripherals following the two-fold modeling methodology, utilize LLMs to find concrete arguments based on the model semantic information, and finally use these arguments to instantiate models to generate peripheral emulators. To this end, we propose \sys to facilitate flexible MCU peripheral emulation as shown in \autoref{fig::design}. The core of \sys uses a predefined set of primitives (parameters) to facilitate the two-fold modeling of various peripherals. Semantic models for each category of peripherals can then be built using these primitives, with each primitive associated with a short description written in natural language to describe its peripheral-related semantics. 
To coordinate this core design, \sys adopts a layered design, inspired by modern compiler infrastructures such as LLVM \cite{Lattner::LLVM:CGO04}, and consists of a frontend and a backend. The frontend takes semantic models as inputs and utilizes an LLM to extract model arguments from peripheral driver source code and generates concrete model instances. These model instances are then sent to the backend, which generates peripheral emulation code through template-based code synthesis. The generated emulation code can be directly integrated into emulators such as QEMU to enable firmware emulation. We elaborate on the details of these components below.



\subsection{Primitives for Two-Fold Modeling}
\label{subsec::design::model_abstraction}
Peripherals, even those of the same category (e.g., UART or DMA), have diverse implementations. Therefore, finding a general abstraction method to model peripherals is challenging. However, through a thorough investigation of MCU peripherals, we have observed that peripherals are structurally composed of a limited set of basic hardware components, with each component assigned different semantics to perform different functionalities. Therefore, we propose a set of 9 primitives for \sys to build abstract models as shown in \autoref{tab::primitive}. These primitives provide high-level abstractions of the three main aspects in peripheral structural characteristics: registers, interrupt events, and memory interactions, as discussed in \S\ref{subsec::background::mcu}.

To further embed functional characteristics into these primitives, a short description must be attached to each primitive to express its semantic. To reduce user burden, \sys has provided descriptions for inner primitives (i.e., primitives used by other primitives), and users only need to describe the outermost primitives that appear in the abstract model. We first elaborate on the 9 primitives, then use an example to illustrate how to build a peripheral semantic model using these primitives.




\noindent\textbf{Register (\texttt{Reg}).} Such primitives correspond to peripheral registers. Each \texttt{Reg} consists of a name, the register width, and the offset within the peripheral.

\noindent\textbf{Register Field (\texttt{RegField}).} Such primitives correspond to fields of peripheral registers. A \texttt{RegField} is associated with a \texttt{Reg} that holds this field, and contains a name, the field width, and the offset within the holding register.

\noindent\textbf{Register Field State (\texttt{RegFieldState}).} The value of a register field can influence the behavior of the peripheral. For example, when the register field responsible for enabling UART transmission is unset, the peripheral should stop transmitting data. Therefore, we introduce this primitive to represent the value of a \texttt{RegField}. A \texttt{RegFieldState} is a pair of a \texttt{RegField} and its value.

\noindent\textbf{Register Field Value Mapping (\texttt{RegFieldMap}).} The value of a register field can be mapped to a different value when it is used by the hardware. For example, DMA peripherals typically use register fields to configure transfer granularity - the hardware transfers data with a 1 byte granularity if the field is set to \texttt{1}, 4 bytes if \texttt{2}, and 8 bytes if \texttt{3}. We introduce this primitive to help model such behaviors. A \texttt{RegFieldMap} primitive consists of a \texttt{RegField} and a mapping from \texttt{RegField} values to mapped values. In the above DMA example, the mapping is $\{1 \to 1, 2 \to 4, 3 \to 8\}$.

\noindent\textbf{Switch (\texttt{Swt}).} Certain hardware functionalities can be enabled/disabled through setting specific register fields. For example, the receiver is enabled/disabled when writing \texttt{1/0} into the \texttt{RE} field of the \texttt{CR1} register in the STM32F4xx \texttt{USART} peripheral. Besides, an additional register field can be used to indicate the current status of this functionality. Therefore, this primitive is introduced to handle such behaviors and consists of three \texttt{RegFieldState} accordingly.

\noindent\textbf{Update (\texttt{Upd}).} When the firmware updates certain register fields (i.e., the condition), other register fields are updated by the hardware (i.e., the action). To express such behaviors, the \texttt{Upd} primitive contains a pair of lists of \texttt{RegFieldState}s, with each representing the condition and action, respectively.

\noindent\textbf{Event (\texttt{Evt}).} Hardware generates signals (typically in the form of setting a register field) when certain events happen. These events are then aggregated to generate interrupt signals. The \texttt{Evt} primitive is hence introduced to help model peripheral events and interrupts. An event is associated with a \texttt{RegFieldState} used as the event status flag, a \texttt{RegFieldState} indicating whether the event can be generated, two \texttt{RegFieldState}s to enable/disable the event, and a \texttt{RegFieldState} to clear the event status flag.
Different events may be used to generate different interrupts (i.e., different interrupt lines).
Therefore, this primitive also contains an interrupt line index to which this event belongs. 

\noindent\textbf{Memory Field (\texttt{MemField}).} While most peripherals are accessed solely through hardware registers, some peripherals can be accessed directly through memory. For example, ENET in the NXP FRDM-K64F MCU is an Ethernet peripheral. It uses in-memory data structures (i.e. transfer descriptors) to manage packet transfers, and the hardware behaviors (e.g., whether the descriptor can be modified by the hardware) can be affected by altering these descriptors. To handle such hardware behaviors, we introduce this primitive to track interesting fields of in-memory data structures. Each \texttt{MemField} contains a field offset and a field width.

\noindent\textbf{Memory Field State (\texttt{MemFieldState}).} The value of a memory field may also influence the behavior of the peripheral as discussed above. Therefore, we introduce this primitive to represent the value of a \texttt{MemField}. A \texttt{MemFieldState} primitive is a pair of a \texttt{MemField} and a value.

\begin{table}[!htb]
\centering
\small
\caption{\textbf{\sys primitives.}}
\label{tab::primitive}
\scriptsize{
\begin{threeparttable}
\begin{tabular}{c c c}
\toprule[1.5pt]
\textbf{Primitives} & \textbf{Members} & \textbf{Member Types} \\
\toprule[1.5pt]
\multirow{3}{*}{\texttt{Reg}} & \texttt{name} & \texttt{string} \\
                              & \texttt{offset} & \texttt{integer} \\
                              & \texttt{width} & \texttt{integer} \\
\midrule
\multirow{4}{*}{\texttt{RegField}} & \texttt{reg} & \texttt{Reg} \\
                                   & \texttt{name} & \texttt{string} \\
                                   & \texttt{offset} & \texttt{integer} \\
                                   & \texttt{width} & \texttt{integer} \\
\midrule
\multirow{2}{*}{\texttt{RegFieldState}} & \texttt{field} & \texttt{RegField} \\
                                        & \texttt{value} & \texttt{integer} \\
\midrule
\multirow{2}{*}{\texttt{RegFieldMap}} & \texttt{field} & \texttt{RegField} \\
                                      & \texttt{map} & \texttt{integer} $\to$ \texttt{integer} \\
\midrule
\multirow{3}{*}{\texttt{Swt}} & \texttt{enable} & \texttt{RegFieldState} \\
                              & \texttt{disable} & \texttt{RegFieldState} \\
                              & \texttt{status} & \texttt{RegFieldState} \\
\midrule
\multirow{2}{*}{\texttt{Upd}} & \texttt{condition} & \texttt{[RegFieldState]} \\
                              & \texttt{action} & \texttt{[RegFieldState]} \\
\midrule
\multirow{6}{*}{\texttt{Evt}} & \texttt{happen} & \texttt{RegFieldState} \\
                              & \texttt{active} & \texttt{RegFieldState} \\
                              & \texttt{enable} & \texttt{RegFieldState} \\
                              & \texttt{disable} & \texttt{RegFieldState} \\
                              & \texttt{clear} & \texttt{RegFieldState} \\
                              & \texttt{irq\_line} & \texttt{integer} \\
\midrule
\multirow{2}{*}{\texttt{MemField}} & \texttt{offset} & \texttt{integer} \\
                                   & \texttt{width} & \texttt{integer} \\
\midrule
\multirow{2}{*}{\texttt{MemFieldState}} & \texttt{field} & \texttt{MemField} \\
                                        & \texttt{value} & \texttt{integer} \\
\bottomrule[1.5pt]
\end{tabular}

\end{threeparttable}
}
\end{table}

Users can easily compose peripheral semantic models using the provided primitives.
Continuing with the timer example in \S\ref{subsec::motivating_example}, we can model the counters of a timer peripheral using two \texttt{Reg} primitives (i.e., tick and period registers), a \texttt{Evt} primitive (i.e., the timer periodic interrupt), and a \texttt{Swt} primitive (i.e., if the counter is enabled), as shown in \autoref{lst::timer-model}. To distinguish their functionalities (e.g., tick register and period register), we then attach a short description written in natural language to express their semantics. For the \texttt{Reg} primitive representing the tick register, we annotate it with \textit{``the register holding the current timer tick value''}. Similarly, for the \texttt{Reg} primitive representing the period register, we annotate it with \textit{``the register holding the timer period value''}. While for the \texttt{Evt} primitive representing the timer periodical interrupt event, we describe it with \textit{``the event generated when the timer tick reaches the period''}. For the \texttt{Swt} primitive, we attach a description \textit{``when to enable the counter''}. Typically, timers also have input capture and output compare functionalities, which can also be similarly modeled as shown in \autoref{lst::timer-model}.

\begin{listing}[H]
\begin{minted}[
breakanywhere,
baselinestretch=1.0,
numbersep=2pt,
bgcolor=bgcolor,
fontsize=\scriptsize,
linenos,
breaklines,
highlightcolor=highlightcolor,
]{python}
class Counter(BaseModel):
  tick: Reg = Field(description="the register holding the current timer tick value")
  period: Reg = Field(description="the register holding the timer period value")
  enable: Swt = Field(description="when to enable the counter")
  period_evt: Evt = Field(description="the event generated when the timer tick reaches the period")

class InputCapture(BaseModel):
  capture: Reg = Field(description="the register holding the capture value")
  enable: Swt = Field(description="when to enable the input capture channel")
  capture_evt: Evt = Field(description="the input capture event")

class OutputCompare(BaseModel):
  compare: Reg = Field(description="the register holding the compare value")
  enable: Swt = Field(description="when to enable the output compare channel")
  compare_evt: Evt = Field(description="the output compare event")

class TimerModel(PeripheralModel):
  counters: list[Counter] = Field(description="a list of counters")
  intput_captures: list[Counter] = Field(description="a list of input capture channels")
  output_compares: list[Counter] = Field(description="a list of output compare channels")

\end{minted}
\small
\caption{\textbf{The complete timer model. \texttt{Field()} is a helper function used to describe the semantic of a primitive.}}
\label{lst::timer-model}
\end{listing}

\subsection{Frontend for Argument Extraction}
\label{subsec::design::frontend}
\sys's frontend extracts concrete model arguments (i.e., primitive instances) from peripheral driver source code, which contains rich peripheral behavior information and has been used to extract peripheral behaviors \cite{Lei::Perry::Sec24}. However, prior solutions depend solely on program analysis techniques and fall short in understanding driver code semantics, thus requiring non-negligible manual effort in understanding and annotating driver code.

To understand driver code semantics and reduce the involved manual efforts, \sys adopts an LLM-empowered approach. For an abstract peripheral model, we first transform the involved primitives as well as their associated descriptions into a \texttt{JSON}-like prompt, which, along with the driver code, is combined with our pre-defined prompt templates (see \refappendix{appendix::prompt-template}) to form the final prompt. This prompt is then sent to the LLM to make it extract model arguments and return the results in a machine-readable format (i.e., \texttt{JSON}). After receiving responses, \sys parses, resolves, and validates the results to form various primitive instances, which are then used to instantiate model instances.

\begin{listing}[!htb]
\begin{minted}[
breakanywhere,
baselinestretch=1.0,
numbersep=2pt,
bgcolor=bgcolor,
fontsize=\scriptsize,
linenos,
breaklines,
highlightcolor=highlightcolor,
]{json}
// JSON-like prompt for the event in the abstract timer model
{
  // the timer periodic event
  "period_evt": {
    // the register field indicating the occurrence of the event
    "happen": {
      "reg": "<register name>",
      "field": "<field name>"
    }
  }
}
\end{minted}

\vspace{-1em}

\begin{minted}[
breakanywhere,
baselinestretch=1.0,
numbersep=2pt,
bgcolor=bgcolor,
fontsize=\scriptsize,
linenos,
breaklines,
highlightcolor=highlightcolor,
highlightlines={2,3,5},
]{c}
// STM32F4xx timer driver code
#define __HAL_TIM_GET_FLAG(HANDLE, FLAG) (((HANDLE)->Instance->SR &(FLAG)) == (FLAG))
#define TIM_FLAG_UPDATE TIM_SR_UIF
/* TIM Update event */
if (__HAL_TIM_GET_FLAG(htim, TIM_FLAG_UPDATE) != RESET) {
  if (__HAL_TIM_GET_IT_SOURCE(htim, TIM_IT_UPDATE) != RESET) {
    __HAL_TIM_CLEAR_IT(htim, TIM_IT_UPDATE);
    // handle the event
    HAL_TIM_PeriodElapsedCallback(htim);
  }
}
\end{minted}

\vspace{-1em}

\begin{minted}[
breakanywhere,
baselinestretch=1.0,
numbersep=2pt,
bgcolor=Green!12,
fontsize=\scriptsize,
linenos,
breaklines,
highlightcolor=highlightcolor,
]{json}
// LLM response for STM32F4xx timers
{
  "period_evt": {
    "happen": {
      "reg": "SR",
      "field": "UIF"
    }
  }
}
\end{minted}
\small
\caption{\textbf{\texttt{JSON}-like prompt for abstract timer model, STM32F4xx timer driver source code, and the extracted arguments by LLM.}}
\label{lst::json-example}
\end{listing}

\begin{sloppypar}
\autoref{lst::json-example} shows how to extract model arguments for the STM32F4xx timer peripheral. \sys first generates a \texttt{JSON}-like prompt for the abstract timer peripheral described in \S\ref{subsec::design::frontend}, with its parameters left as blanks. For simplicity, the shown prompt only contains the \texttt{happen} member for the periodical $Evt$. This prompt, combined with the corresponding timer driver code, is then sent to the LLM. As can be seen from the code, the driver first checks if there is a periodical event by checking if the \texttt{UIF} field of the \texttt{SR} register is set (line 4), then handles the event in line 12. Surprisingly, the LLM understands the code snippet well and returns a \texttt{JSON} object with the above information filled. However, three challenges (\textbf{C}) prevent the automation of the above process due to the limitations of LLMs:
\end{sloppypar}

\noindent\textbf{(\textbf{C-I}) Longer Inputs, Worse Results.} An abstract model can consist of multiple primitives, with each primitive containing other primitives and basic elements. This implies a long prompt, which inevitably increases the LLM input length. However, we have observed a reasoning performance drop as the input length increases, which is also validated by recent work \cite{Levy::LLMInputLen::ACL24}. To make things worse, most LLMs only allow a relatively small number of output tokens (e.g., 4096 or 8192). This makes it impossible to extract all the information described in the prompt generated for an abstract peripheral model. Therefore, it is essential to reduce input length.

\noindent\textbf{\textbf{(C-II)} Limited Instruction Following Ability.} LLMs may discard instructions and generate unexpected results when extracting model arguments, e.g., returning a string instead of an integer value. Different from hallucinations (which we discuss later), such mistakes are typically benign - the results are still logically correct but in an unexpected format. For example, when we ask the LLM to return an integer value, it may instead return the name of a macro representing the expected value. Therefore, we need to identify such scenarios and resolve the expected results from the unexpected ones.

\noindent\textbf{(\textbf{C-III}) Hallucinations.} The most criticized flaw of LLMs is their inevitable tendency to hallucinate, which may result in wrong model arguments and affect the emulation of peripherals. To improve the quality of the generated emulators, we must detect LLM hallucinations and reject wrong responses.

We propose three solutions (\textbf{S}) to handle the above challenges. To handle \textbf{C-I}, \sys breaks down the overall task into multiple smaller sub-tasks and performs multi-stage queries to keep a reasonably short input length. Regarding \textbf{C-II}, \sys adopts a code analysis-based scheme to resolve unexpected responses into expected ones. For \textbf{C-III}, \sys detects LLM hallucinations by detecting self-contradictions among responses. We elaborate on these procedures below.


\noindent\textbf{(\textbf{S-I}) Task Decomposition.} We break the overall argument extraction task into 7 sequentially dependent sub-tasks (which we call stages). For each stage, we provide driver source code to the LLM, and additionally send a simple and single-minded prompt to ask the LLM to extract part of the model arguments. Based on LLM responses, a stage further generates a collection of currently extracted arguments by combining results from previous stages. The 7 stages are listed below:

\begin{figure*}[!htb]
\centering
\begin{minipage}{.37\linewidth}
\begin{minted}[
breakanywhere,
baselinestretch=1.0,
numbersep=2pt,
bgcolor=bgcolor,
fontsize=\scriptsize,
breaklines,
highlightcolor=highlightcolor,
highlightlines={5,6,8},
]{json}
// LLM response
{
  "name":"RTC",
  "irqs":[
    "RTC_WKUP_IRQn",
    "RTC_Alarm_IRQn"
  ],
  "base":"APB1PERIPH_BASE + 0x00002800UL",
}
\end{minted}
\end{minipage}
\begin{minipage}{.52\linewidth}
\begin{minted}[
breakanywhere,
baselinestretch=1.0,
numbersep=2pt,
bgcolor=bgcolor,
fontsize=\scriptsize,
breaklines,
highlightcolor=highlightcolor,
highlightlines={6,7,10,11},
]{c}
// driver code
#define PERIPH_BASE (0x40000000UL)
#define APB1PERIPH_BASE PERIPH_BASE

typedef enum {
  RTC_WKUP_IRQn = 3,
  RTC_Alarm_IRQn = 41,
} IRQn_Type;

// added by source resolver
u64 SOURCE_RESOLVER_MARKER = APB1PERIPH_BASE + 0x00002800UL;
\end{minted}
\end{minipage}
\begin{minipage}{.01\linewidth}
\begin{tikzpicture}[overlay, remember picture]
\draw[->,thick,ForestGreen] (-12.7, -0.29) to[in=180, out=0] (-9, -0.3);
\draw[->,thick,ForestGreen] (-12.7, -0.02) to[in=180, out=0] (-9, 0);
\draw[->,thick,ForestGreen] (-12.1, -0.98) to[in=180, out=320] (-9.24, -1.3);
\end{tikzpicture}
\end{minipage}
\small
\vspace{-1em}
\caption{Example of code analysis based resolution.}
\label{fig::example-resolution}
\end{figure*}

\begin{packeditemize}
\item[\circlenum{1}\textbf{Peripheral Category Identification}]: Based on existing peripheral models, we ask the LLM to identify all peripheral categories within the target MCU and associate them with an available peripheral model. The output is a map from peripheral category names to model names. For example, assume that we provide \sys with two peripheral models -- \texttt{UART} and \texttt{Timer}, and that the drivers have peripherals \texttt{USART} and \texttt{TIM}, the output should be \texttt{\{``USART'': ``UART'', ``TIM'': ``Timer''\}}.
\item[\circlenum{2}\textbf{Register Identification}]: For each peripheral category, we ask the LLM to identify all registers within the peripheral and return a list of \texttt{Reg} primitive instances. Each \texttt{Reg} primitive instance should contain members listed in \autoref{tab::primitive}. For example, the response should look like \texttt{\{[``name'': ``SR'', ``offset'': ``32'', ``width'': ``32''], ...\}}.
\item[\circlenum{3}\textbf{Register Field Identification}]: For each identified register, we ask the LLM to identify all fields within it and return a list of \texttt{RegField} primitive instances. These results are then aggregated to form a mapping from \texttt{Reg}s to their associated \texttt{RegField}s.
\item[\circlenum{4}\textbf{Update Identification}]: For each peripheral category, we ask the LLM to identify register update dependencies, and return a list of \texttt{Upd} primitive instances.
\item[\circlenum{5}\textbf{Peripheral Semantic Identification}]: The above stages focus on extracting peripheral-independent information, while this stage extracts peripheral-specific semantics. To achieve this, this stage asks the LLM to extract all arguments that are not extracted by stage \circlenum{1}-\circlenum{3}. Note that since stage \circlenum{1} already extracted all \texttt{Reg} and \texttt{RegField} primitive instances, we only ask the LLM to extract the names of \texttt{Reg} and \texttt{RegField} instances, thus further reducing the prompt length.
\item[\circlenum{6}\textbf{Peripheral Instance Identification}]: A peripheral category may have multiple instances within the target MCU, each mapped at a different physical address. For example, STM32F4xx MCUs have 4 USART instances mapped at different addresses (i.e., \texttt{USART1}, \texttt{USART2}, \texttt{USART3}, \texttt{USART6}), but all of them are of the \texttt{USART} category. We ask the LLM to identify all instances for each peripheral category, and output a list of instance names, base addresses, and interrupt line numbers. For the above example, the output should be: \texttt{[\{``instance'': ``USART1'', ``base'': ``0x40011000'', ``irqs'': [37]\}, \{``instance'': ``USART2'', ``base'': ``0x40004400'', ``irqs'': [38]\}, ...]}.
\item[\circlenum{7}\textbf{Interrupt Association}]: For each peripheral instance, we ask the LLM to associate each contained \texttt{Evt} primitive with the actual interrupt line number. The identified line number is then stored back into the \texttt{irq\_line} member of an \texttt{Evt} instance.
\end{packeditemize}

\noindent\textbf{(\textbf{S-II}) Resolution by Code Analysis.} LLMs may output unexpected results when we try to extract certain integer values from driver code. These results are logically correct, but cannot be parsed and understood by the machine. For example, when extracting information about a \texttt{RegFieldState} primitive, the LLM may return a string (e.g., macro or enum names, expressions) as its value, although the value is specified to be an integer in the \texttt{JSON}-like prompt, thus causing a parsing error. However, we find that unexpected results typically refer to names of units in the driver code. Following this observation, we propose a code analysis-based resolution scheme to handle unexpected results. Every time \sys fails to directly parse a string into an integer, it first treats the string as the name of a global constant (global variables and enum constants), and searches for the corresponding global constant in the driver code. If \sys gets a match, the concrete value of the identified constant is used. Otherwise, \sys treats the string as a constant expression (e.g., \texttt{SOME\_MACRO << 1}), and constructs a program where the expression is assigned to a variable. If the program compiles and \sys successfully gets the concrete value of the constructed variable, the obtained value is used as the resolution result. If the resolution process fails, \sys discards this response and asks the LLM to generate a new response.

\autoref{fig::example-resolution} shows an example of the resolution process. The LLM returns strings (names and expressions) as the interrupt line numbers and base address of the RTC peripheral in the peripheral instance identification stage, which cannot be directly parsed by the machine. To handle such situations, \sys first finds in the source code for global constants named by these strings, and \texttt{RTC\_WKUP\_IRQn} and \texttt{RTC\_Alarm\_IRQn} are successfully resolved. The remaining string is a constant expression involving macros. Therefore, \sys introduces a global variable initialized by this expression, and the value of this expression can be resolved by inspecting the initial value of the global variable.

\noindent\textbf{(\textbf{S-III}) Validation by Self-Contradiction.} The extracted model arguments can be wrong due to LLM hallucinations. However, we observe that LLM responses are generally reliable when they are self-consistent. For example, there should be no address overlapping between identified registers. Based on this observation, we filter out invalid responses by detecting self-contradictions in the results within a single stage and across multiple stages, and re-run the corresponding stages until we receive a valid response. 
We adopt different detection strategies for different stages. For stage \circlenum{1}, we check for repeated peripheral category names. For stage \circlenum{2}, we check for overlapping registers: for two registers $Reg_a$ and $Reg_b$, they overlap when $\left[Reg_a.offset, Reg_a.offset + Reg_a.width\right)$ intersects with $\left[Reg_b.offset, Reg_b.offset + Reg_b.width\right)$. For stage \circlenum{3}, we check for overlapping fields: for two register fields $RegField_a$ and $RegField_b$, they overlap when

{
\vspace{-1em}
\footnotesize
\begin{align*}
&\left[RegField_a.offset, RegField_a.offset + RegField_a.width\right)\\
\cap&\left[RegField_b.offset, RegField_b.offset + RegField_b.width\right)
\end{align*}
\vspace{-2em}
}

\noindent results in an non-empty set. For stage \circlenum{4} and \circlenum{5}, we make sure every returned register and register field name can be found within the results from stage \circlenum{2} and \circlenum{3}. For stage \circlenum{6}, we check for repeated peripheral instance names, base addresses, and interrupt line numbers. For stage \circlenum{7}, we check whether the identified interrupt line numbers also appear in the results of stage \circlenum{6}. If the validation process fails for responses obtained from one stage, these results are discarded, and this stage as well as all dependent stages are re-executed until there are no invalid responses.

\subsection{Backend for Emulator Generation}
\label{subsec::design::backend}
\sys's backend is responsible for generating the actual emulator code for peripheral model instances. Following the design of primitive-driven abstract models, \sys's backend adopts a template-based code synthesis approach, with some implementation details (i.e., primitives) omitted as blanks. These blanks are later filled by primitive instances inferred by the frontend to form a valid emulator. Therefore, the backend can generate dedicated emulation logic for different model instances.

The overall synthesis procedure can be divided into peripheral-independent and peripheral-specific parts. In the peripheral-specific part, users need to write the synthesis logic to emulate different peripheral behaviors. Although this requires manual effort, it is worth noting that it is a one-time task - the synthesis logic can be reused to generate various peripheral emulators once it is completed. Continuing with the timer example in \S\ref{subsec::motivating_example}, the peripheral-specific part generates the following emulator logic for timer peripherals: the value within the tick \texttt{Reg} is increased by 1 periodically; once it reaches the value within the period \texttt{Reg}, a timer \texttt{Evt} happens, and the event status flag is set according to the corresponding \texttt{RegFieldState} primitive.

To further reduce the required manual efforts in implementing peripheral-specific synthesis logic, \sys also provides built-in peripheral-independent synthesis logic. For \texttt{Upd} primitives, \sys generates code to emulate register update dependencies. Specifically, \sys sets the action register field according to the action \texttt{RegFieldState} when the condition register field is updated to the value specified in the condition \texttt{RegFieldState}. \sys also generates code to emulate event-driven interrupts. To achieve this, \sys first groups \texttt{Evt} primitives based on interrupt line numbers. For each group, \sys first generates an interrupt firing function, where a conjunction of all status flags and enable flags is calculated to set the current level of the interrupt line. Then, \sys injects a call to this function every time register fields used to enable/disable/clear these \texttt{Evt}s are updated. In this way, interrupts will be raised when certain events happen, and lowered when events are disabled or cleared. Besides these, \sys also provides basic logic to manipulate register values and field values to further reduce the implementation cost of peripheral-specific synthesis logic.

\section{Implementation}
\label{sec::impl}

We have implemented a prototype of \sys with $\sim8,000$ lines of Python and C++ code. Most components are written in Python, except for the code analysis-based resolution scheme (\S\ref{subsec::design::frontend}), which is built upon LibTooling \cite{LLVM::LibTooling::Web} and LLVM \cite{Lattner::LLVM:CGO04} and is written in C++. Regarding LLM integration, \sys now supports Google's Gemini models \cite{Google::Gemini::Arxiv} because Google has publicly provided free access to the Gemini API. At the backend, \sys currently supports generating emulators for QEMU \cite{Bellard::QEMU::ATC05}, i.e., the generated emulators can be directly integrated into QEMU. \sys now supports ARM Cortex-M, which is the dominant architecture and is used by $\geq70\%$ MCUs according to a recent investigation \cite{Nino::IOT::SEC24}. However, the design of \sys does not target specific architectures and extending \sys to support other architectures only requires engineering efforts, which we leave for future work.

The code analysis based resolution scheme (see \S\ref{subsec::design::frontend}) first builds Abstract Syntax Trees (AST) for the target driver library, which requires a compilation database \cite{LLVM::CompDB::Web} as the input. This can be easily obtained by compiling the library using Bear \cite{rizsotto::Bear::Web} or CMake \cite{kitware::CMake::Web}. Then, \sys utilizes LibTooling to traverse the driver's ASTs, and searches for matching global constants during the traversal. Due to the existence of macros, when resolving an expression, \sys constructs a program that includes all driver headers, and a global variable \texttt{SOURCE\_RESOLVER\_MARKER} (see \autoref{fig::example-resolution}) of integer type initialized by the target expression. This program is compiled into LLVM bitcode using LLVM APIs, and \sys searches in the resulting LLVM bitcode for the introduced global variable. Once \sys gets a match, the constant value used to initialize this variable is extracted as the resolved integer value.

\sys has a modular design and can be easily extended to model various peripherals. We have successfully applied \sys to model 12 kinds of commonly used MCU peripherals: analog-to-digital converter (ADC), digital-to-analog converter (DAC), DMA controller, Ethernet controller, general purpose input/output (GPIO), random number generator (RNG), secure digital high capacity (SDHC), secure digital input output (SDIO), serial peripheral interface (SPI), inter-integrated circuit (I2C), timers, and universal asynchronous receiver-transmitter (UART). For other peripherals, \sys applies a basic model to support basic register operations and register update dependencies. We provide the model for DMA controllers as an example in \refappendix{appendix::dma-example}.

\section{Evaluation}
\label{sec::eval}
In this section, we conduct evaluation experiments to answer the following research questions (\textbf{RQ}):

\begin{packeditemize}
\item[\textbf{(RQ 1):}] Can emulators generated by \sys perform consistently with actual hardware?
\item[\textbf{(RQ 2):}] Can emulators generated by \sys be used to rehost various firmware?
\item[\textbf{(RQ 3):}] Can \sys be easily extended to support more kinds of peripherals?
\item[\textbf{(RQ 4):}] What benefits can \sys bring to firmware security?
\end{packeditemize}

\begin{table}[htb!]
    \centering
    \setlength{\belowrulesep}{2pt}
    \setlength{\aboverulesep}{2pt}
    \setlength{\tabcolsep}{5pt}
    \small
    \caption{\textbf{Summary of target MCUs}}
    \label{tab::mcu-targets}
    \scriptsize{
        \begin{tabular}{c l r}
            \toprule[1.5pt]
            \textbf{Vendor} & \textbf{MCUs} & \textbf{Driver Libraries} \\
            \toprule[1.5pt]
            \multirow{6}{*}{ST}         & STM32F072 & STM32CubeF0 v1.11.3  \\
                                        & STM32F103 & STM32CubeF1 v1.8.4   \\
                                        & STM32F407 & STM32CubeF4 v1.26.2  \\
                                        & STM32F429 & STM32CubeF4 v1.26.2  \\
                                        & STM32F769 & STM32CubeF7 v1.16.2  \\
                                        & STM32L073 & STM32CubeL0 v1.12.1  \\
            \midrule
            \multirow{4}{*}{NXP}        & FRDM-K22F & MCUXpresso SDK v2.12.0     \\
                                        & FRDM-K64F & MCUXpresso SDK v2.11.0     \\
                                        & FRDM-K82F & MCUXpresso SDK v2.8.0      \\
                                        & FRDM-KL25Z & MCUXpresso SDK v2.2.0      \\
            \midrule
            \multirow{5}{*}{Microchip}  &    SAM4L-EK & \multirow{5}{*}{\begin{tabular}[c]{@{}c@{}}Advanced Software Framework v3.52.0\end{tabular}} \\
                                        &   SAM4E Xplained Pro & \\
                                        &   SAM4S Xplained & \\
                                        &   SAM E70 Xplained & \\
                                        &   SAM V71 Xplained Ultra & \\
                                        &   SAM3X8E & \\
            \bottomrule[1.5pt]
        \end{tabular}
    }
\vspace{-1em}
\end{table}

\subsection{Experiment Setup}
\label{subsec::experiment-setup}
We use a dataset consisting of 90 firmware samples to evaluate \sys: 46 firmware samples from P2IM unit tests, 10 real-world firmware samples from P2IM, 7 Ethernet firmware samples from HALucinator, 4 DMA firmware samples from DICE, 19 shell firmware samples and a BLE host firmware from Perry, and 3 Bluetooth host stack firmware from Zephyr \cite{Zephyr::Zephyr::Web}, NuttX \cite{NuttX::NuttX::Web}, and Mynewt/NimBLE \cite{Mynewt::Mynewt::Web}. These firmware samples involve 15 MCUs manufactured by 3 leading vendors and cover a wide range of peripherals, including complex peripherals like DMA. We first use \sys to generate emulators for these MCUs, then evaluate these emulators with the above firmware samples. \autoref{tab::mcu-targets} lists the involved MCUs and the used driver libraries (see \S\ref{subsec::design::frontend}). Regarding the LLM to be used by \sys, we choose the Gemini 1.5 Flash model due to its higher rate limit. The model temperature is set to 1.0 by default.


\begin{table*}[tbh!]
\centering
\setlength{\belowrulesep}{2pt}
\setlength{\aboverulesep}{2pt}
\setlength{\tabcolsep}{8pt}
\small
\caption{\textbf{Evaluation results on P2IM unit tests.}}
\label{tab::p2im-unit-tests}
\scriptsize{
\begin{threeparttable}
\begin{tabular}{l l c c c c c c || r}
\toprule[1.5pt]
\multirow{2}{*}{\textbf{Peri.}} & \multirow{2}{*}{\textbf{Unit Test}} & \multicolumn{3}{c}{\textbf{STM32F103}} & \textbf{FRDM-K64F} & \multicolumn{2}{c||}{\textbf{ATSAM3X8E}} & \tabincell{c}{\textbf{Passing}\\\textbf{Rate}} \\
& & Arduino & RIOT* & NUTTX & RIOT & Arduino & RIOT & \\
\toprule[1.5pt]
ADC                     & read converted values         & \cmark    & -         & \cmark    & \xmark (0/3)    & \cmark & \cmark & 4/5 \\
DAC                     & write values for conversion   & -         & -         & -         & -         & \cmark & \cmark & 2/2 \\
\multirow{3}{*}{GPIO}   & execute the interrupt callback& \cmark    & \cmark    & \cmark    & \cmark    & \cmark & \cmark & 6/6 \\
                        & read a pin                    & \cmark    & \cmark    & \cmark    & \cmark    & \cmark & \cmark & 6/6 \\
                        & set/clear a pin               & \cmark    & \cmark    & \cmark    & \cmark    & \cmark & \cmark & 6/6 \\
PWM                     & perform basic configuration   & \cmark    & -         & \cmark    & \cmark    & \cmark & \cmark & 5/5 \\
\multirow{2}{*}{I2C}    & receive bytes                 & \cmark    & -         & \cmark    & \cmark    & \cmark & -      & 4/4 \\
                        & send bytes                    & \cmark    & -         & -         & \cmark    & \cmark & -      & 3/3 \\
\multirow{2}{*}{UART}   & receive bytes                 & \cmark    & \cmark    & \cmark    & \cmark    & \cmark & \cmark & 6/6 \\
                        & transmit bytes                & \cmark    & \cmark    & \cmark    & \cmark    & \cmark & \cmark & 6/6 \\
\multirow{2}{*}{SPI}    & receive bytes                 & \cmark    & \cmark    & \cmark    & \cmark    & \cmark & \cmark & 6/6 \\
                        & transmit bytes                & \cmark    & \cmark    & -         & \cmark    & \cmark & \cmark & 5/5 \\
\multirow{2}{*}{TIMER}  & execute the interrupt callback& -         & \cmark    & -         & \cmark    & -      & \cmark & 3/3 \\
                        & read counter values           & -         & \cmark    & -         & \cmark    & -      & \cmark & 3/3 \\
\bottomrule[1.5pt]
     & & \multicolumn{3}{c}{} &  & \multicolumn{2}{c||}{} & \textbf{65/66(98.48\%)}\\
\bottomrule[1.5pt]
\end{tabular}

\begin{tablenotes}
\footnotesize
\item[] - : No such unit test. \cmark: All the 3 generated emulators pass this unit test. \xmark($i$/3): $i$ (out of 3) generated emulators pass this unit test.
\end{tablenotes}
\end{threeparttable}
}
\end{table*}

We first evaluate the passing rate achieved by the generated emulators on P2IM unit tests to answer \textbf{RQ 1}. Then, we use these emulators to rehost the remaining firmware samples to examine whether they can achieve expected functionalities, and evaluate the success rate to answer \textbf{RQ 2}. To answer \textbf{RQ 3}, we evaluate the required lines of code to develop an abstract peripheral model and its corresponding backend to validate \sys's extensibility. Finally, we use the generated emulators to fuzz the Bluetooth host stacks of three popular RTOSes to answer \textbf{RQ 4}.

To understand \sys's improvements in firmware-agnostic emulation, we compare \sys with state-of-the-art solutions, Perry \cite{Lei::Perry::Sec24} and SEmu \cite{Zhou::sEmu::CCS22}. Since Perry also uses driver source code to generate emulators, we use the same driver source code with Perry in the evaluation for a fair comparison. All experiments were conducted on a machine featuring an Intel Core™ i7-1165G7 @ 2.80GHz processor, running Ubuntu 20.04 LTS, and equipped with 16GB of RAM.

\subsection{Emulator Consistency}
\label{subsec::consistency}
Following prior work \cite{Zhou::sEmu::CCS22, Lei::Perry::Sec24}, we use P2IM \cite{Feng::P2IM::Sec20} unit tests to validate whether the generated emulators can replicate hardware behaviors (\textbf{RQ 1}). P2IM provides 46 firmware samples containing 66 valid unit tests, covering 8 peripheral categories on different combinations of 3 MCUs (STM32F103, FRDM-K64F and ATSAM3X8E) and 3 OS libraries (Arduino, RIOT and NuttX). We use \sys to generate emulators for the 3 MCUs, rehost each firmware sample, and compare the recorded execution trace with the ground truth, i.e., the execution trace when the firmware is executed on the hardware. To minimize the impact of random factors introduced by the LLM, we repeat each generation process 3 times to generate 3 variant emulators of a MCU. We evaluate the passing rate on these unit tests using the generated MCUs, and the results are shown in \autoref{tab::p2im-unit-tests}. Without manual intervention, \sys passes 98.48\% (65/66) of the unit tests, surpassing Perry and SEmu, whose passing rates are 74.24\% and 0\%, respectively.

The improvements brought by \sys mainly attribute to two aspects - the abstract modeling of peripherals and the usage of LLMs. \sys builds different abstract models for different categories of peripherals to capture peripheral-specific functionalities, making the generated emulators more faithful. On the contrary, SEmu and Perry try to model all hardware functionalities with a small set of heuristic rules, which is infeasible due to diverse peripheral functionalities. For example, SEmu fails to emulate I2C peripherals because it does not know that the least significant bit of the device address encodes transfer directions, which is defined in the I2C specification \cite{NXP::I2CSpec::web} and followed by various I2C implementations. However, \sys embeds such knowledge into its I2C abstract models and can hence generate faithful I2C peripheral emulators. On the other hand, the usage of LLMs has equipped \sys with a strong semantic understanding capability to understand the used peripheral description materials and extract correct arguments. For example, Perry fails to emulate interrupt events for ATSAM3X8E UART using program analysis techniques - it cannot infer that setting specific fields of the \texttt{IER}/\texttt{IDR} register will set/clear corresponding fields in the \texttt{IMR} register, which controls whether an interrupt event can be generated. However, \sys successfully captures such functionalities because LLMs can understand semantics conveyed through natural language (e.g., function names). \looseness=-1

Although we enhance \sys with capabilities to detect logically incorrect LLM responses (see \S\ref{subsec::design::frontend}) to make the pipeline work, \sys may still accept wrong answers, which causes the failing cases in \autoref{tab::p2im-unit-tests}. Specifically, the LLM returns a correct but incomplete \texttt{Upd} primitive instance for the ADC model, with an action omitted. As a consequence, the generated emulator fails to set a register field that the drivers infinitely poll for, eventually leading to program hang. However, such errors rarely happen and fixing this error requires minimal manual efforts -- one only needs to add one line of code (LoC) to the generated emulator. As a comparison, Perry needs 6 LoC to fix wrong models, while SEmu needs to fix and add 16 generated rules.


\begin{table}[tbh!]
    \centering
    \setlength{\belowrulesep}{2pt}
    \setlength{\aboverulesep}{2pt}
    \setlength{\tabcolsep}{4pt}
    \small
    \caption{\textbf{Firmware emulation results.}}
    \label{tab::universality}
    \scriptsize{
        \begin{threeparttable}
            \begin{tabular}{l l l l c}
                \toprule[1.5pt]
                \textbf{MCUs} & \textbf{Firmware} & \tabincell{c}{\textbf{\# Miss.}\\\textbf{Func.}} & \tabincell{c}{\textbf{\# Wrong}\\\textbf{Func.}} \\
                \midrule[1.5pt]
                \multirow{2}{*}{STM32F072}              & Zephyr-Shell      & \multirow{2}{*}{1 (\ding{73})} & \multirow{2}{*}{0} \\
                                                        & LiteOS-Shell      & & \\
                \midrule
                \multirow{10}{*}{STM32F103}              & Zephyr-Shell      & \multirow{10}{*}{0} & \multirow{10}{*}{0} \\
                                                        & LiteOS-Shell      & & \\
                                                        & Drone             & & \\
                                                        & Gateway           & & \\
                                                        & Reflow\_Oven      & & \\
                                                        & Robot             & & \\
                                                        & Soldering\_Iron   & & \\
                                                        & Soldering Station & & \\
                                                        & GPS Receiver      & & \\
                                                        & Oscilloscope      & & \\
                \midrule
                \multirow{3}{*}{STM32F407}              & Zephyr-Shell      & \multirow{2}{*}{0} & \multirow{2}{*}{0} \\
                                                        & Zephyr-BLE-Host   & & \\
                                                        & NuttX BLE Host    & & \\
                \midrule
                \multirow{10}{*}{STM32F429}              & LiteOS-Shell      & \multirow{8}{*}{0} & \multirow{8}{*}{0} \\
                                                        & CNC               & & \\
                                                        & PLC               & & \\
                                                        & MIDI Synthesizer  & & \\
                                                        & UDP Echo Client   & & \\
                                                        & UDP Echo Server   & & \\
                                                        & TCP Echo Client   & & \\
                                                        & TCP Echo Server   & & \\
                                                        & Zephyr BLE Mesh   & & \\
                                                        & Mynewt/NimBLE Mesh   & & \\
                \midrule
                \multirow{2}{*}{STM32F769}              & Zephyr-Shell      & \multirow{2}{*}{0} & \multirow{2}{*}{0} \\
                                                        & LiteOS-Shell      & & \\
                \midrule
                \multirow{2}{*}{STM32L073}              & Zephyr-Shell      & \multirow{2}{*}{1 (\ding{73})} & \multirow{2}{*}{0} \\
                                                        & LiteOS-Shell      & & \\
                \midrule[1.5pt]
                FRDM-K22F                               & Zephyr-Shell      & 0 & 0 \\
                \midrule
                \multirow{5}{*}{FRDM-K64F}              & Zephyr-Shell      & \multirow{5}{*}{0} & \multirow{5}{*}{0} \\
                                                        & Console           & & \\
                                                        & HTTP Server       & & \\
                                                        & TCP Echo Server   & & \\
                                                        & UDP Echo Server   & & \\
                \midrule
                FRDM-K82F                               & Zephyr-Shell      & 0 & 0 \\
                \midrule
                FRDM-KL25Z                              & Zephyr-Shell      & 0 & 0 \\
                \midrule[1.5pt]
                SAM4L-EK                                & Zephyr-Shell      & 0 & 0 \\
                \midrule
                SAM4E Xplained Pro                      & Zephyr-Shell      & 0 & 0 \\
                \midrule
                SAM4S Xplained                          & Zephyr-Shell      & 0 & 0 \\
                \midrule
                SAM E70 Xplained                        & Zephyr-Shell      & 0 & 0 \\
                \midrule
                SAM V71 Xplained Ultra                  & Zephyr-Shell      & 0 & 0 \\
                \midrule
                \multirow{2}{*}{SAM3X8E}                & Heat\_Press       & \multirow{2}{*}{0} & \multirow{2}{*}{0} \\
                                                        & Steering\_Control & & \\
               \bottomrule[1.5pt]
            \end{tabular}

            \begin{tablenotes}
                \footnotesize
                \item[\textbf{Note}] \ding{73}: Unmodeled hardware functionality.
            \end{tablenotes}
        \end{threeparttable}
    }
\end{table}

\subsection{Emulator Universality}
\label{subsec::universality}
We demonstrate the universality (\textbf{RQ 2}) of the generated emulators by rehosting 44 firmware samples: 10 real-world firmware from P2IM \cite{Feng::P2IM::Sec20}, 7 Ethernet firmware from HALucinator \cite{Clements::HALucinator::Sec20}, 4 real-world DMA firmware from DICE \cite{Mera::DICE::SP20}, 19 shell firmware and one BLE host firmware from Perry \cite{Lei::Perry::Sec24}, and 3 Bluetooth host firmware from Zephyr \cite{Zephyr::Zephyr::Web}, NuttX \cite{NuttX::NuttX::Web} and Mynewt/NimBLE \cite{Mynewt::Mynewt::Web}, respectively. Although the shell firmware samples seem trivial to rehost, it is important to note that their initialization logic involves extensive hardware interactions (e.g., clock configurations), which Perry and SEmu struggle to handle. We have manually analyzed involved firmware samples and confirmed that they all use peripherals to implement expected functionalities. A firmware sample is successfully rehosted if it can achieve corresponding functionalities as defined in its code logic. Since firmware functionalities directly depend on hardware functionalities provided by peripherals, the above approach can effectively validate the functionalities of emulated peripherals. For example, the HTTP Server firmware uses the ETH peripheral to send and receive Ethernet frames. Therefore, if we send an HTTP request to the rehosted HTTP Server firmware and it returns the hard-coded web page, the emulated ETH peripheral must be working normally.

For firmware that cannot be properly rehosted by the generated emulator, we inspect the involved emulation code and record the number of involved wrong or missing hardware functionalities. The evaluation results are shown in \autoref{tab::universality}. Without manual intervention, \sys achieves a 95.45\% (42/44) rehosting success rate, while Perry and SEmu only successfully rehost 23 and 0 firmware. The only exceptions occur when rehosting the LiteOS-Shell firmware for STM32F072 and STM32L073. The SYSCFG peripheral of these two MCUs uses a register to configure the base address of the CPU's interrupt vector table, which is a rarely seen functionality and is not initially covered by the 12 supported peripheral models. However, we are able to use \sys to build a dedicated model for this peripheral with only one \texttt{RegFieldMap} primitive, which maps different register field values into different base addresses. This also illustrates the extensibility of \sys and answers \textbf{RQ 3}.

\subsection{Framework Extensibility}
\label{subsec::extensibility}
\sys's extensibility is reflected in two aspects. First, a peripheral model can generate multiple emulators for different peripherals. Second, supporting a new kind of peripheral in \sys requires approximately the same engineering effort as developing a dedicated peripheral emulator. We then illustrate the two aspects and answer \textbf{RQ 3}.

The first aspect has been validated in \S\ref{subsec::consistency} and \S\ref{subsec::universality}, where we use the peripheral models to generate emulators for different peripherals of different MCUs. Additionally, we conduct a case study to further illustrate this point. Ethernet controllers are considered complex peripherals. They use transfer descriptors to manage the reception and transmission of Ethernet frames. Different hardware implementations have distinct transfer descriptor layouts. Both \sys and SEmu \cite{Zhou::sEmu::CCS22} are capable of emulating Ethernet controllers. However, we find that SEmu only supports STM32 Ethernet controllers because their model is highly coupled with the specific transfer descriptor layout of this hardware implementation. Therefore, it is almost impossible to extend their model to support other Ethernet controllers like the one used by NXP FRDM-K64F. However, with the help of \sys's two-fold modeling mechanism, we successfully build a generic peripheral model (see \refappendix{appendix::eth}) for Ethernet controllers. This model covers common behaviors of Ethernet controllers and supports both STM32 and the NXP FRDM-K64F Ethernet controllers, as shown in \autoref{tab::universality}.

\begin{table}[htb!]
    \centering
    \setlength{\belowrulesep}{2pt}
    \setlength{\aboverulesep}{2pt}
    \setlength{\tabcolsep}{5pt}
    \small
    \caption{\textbf{Required engineering efforts to implement \sys backends and QEMU emulators.}}
    \label{tab::extensibility}
    \scriptsize{
        \begin{tabular}{r r r r}
            \toprule[1.5pt]
            \textbf{Model} & \textbf{LoC (\sys)} & \tabincell{c}{\textbf{LoC (QEMU) /}\\\textbf{\# Peripherals}} & \textbf{LoC Diff w.r.t. QEMU}\\
            \toprule[1.5pt]
            ADC         & 256       & 397 / 1 & -141 \\
            DAC         & 223       & - / 0   &  / \\
            DMA         & 399       & - / 0   &  / \\
            Ethernet    & 653       & - / 0   &  /  \\
            GPIO        & 207       & 548 / 1 & -341 \\
            I2C         & 446       & - / 0   & /  \\
            RNG         & 276       & - / 0   & /  \\
            SDHC        & 298       & - / 0   & /  \\
            SDIO        & 351       & - / 0   & /  \\
            SPI         & 238       & 297 / 1 & -59  \\
            Timer       & 617       & 449 / 1 & +168  \\
            UART        & 371       & 328 / 1 & +43  \\
            \midrule[1.5pt]
            \textbf{Avg.} & 361.25 & 403.8 & -66.0 \\
            \bottomrule[1.5pt]
        \end{tabular}
    }
\end{table}
In \sys, the main engineering efforts to support a kind of peripherals concentrate on implementing the backend (i.e., peripheral emulator templates).
To validate the second aspect, we collect the required lines of code (LoC) to develop backends for the 12 supported peripheral models in \sys. For comparison, we also collect manually implemented peripheral emulators in QEMU (commit \texttt{853546}) for MCUs listed in \autoref{tab::mcu-targets} and count the required LoC in total. The results are listed in \autoref{tab::extensibility}. Most peripherals of the target MCUs are not supported by QEMU and we only find 5 emulated peripherals, with each peripheral category contains at most one peripheral emulator, which highlights the need for automated peripheral emulation. On average, it takes  $\sim$361 LoC to implement a peripheral model backend (i.e., QEMU peripheral emulator templates) in \sys, and $\sim$403 LoC to implement a peripheral emulator in QEMU. Considering that implementing the backend is similar to implementing an actual QEMU peripheral emulator, comparing LoC of \sys' backends with QEMU's peripheral emulators can effectively reflect the required engineering efforts. Therefore, we can conclude that supporting a kind of peripherals in \sys requires approximately the same engineering efforts than developing a dedicated peripheral emulator in QEMU. Although it may take more LoC to implement the backends under some cases (e.g., UART and Timer), we argue that the extra efforts are worthwhile because the backend can derive multiple emulators. Besides, due to \sys's modular design, users can even reuse some of the emulator code generation logic for different peripheral model backends. For example, the SPI model and the UART model have similar primitives because they both send and receive data through data registers and have similar interrupt events. Therefore, their backends share some of the emulator generation code.

\begin{table}[htb!]
    \centering
    \setlength{\belowrulesep}{2pt}
    \setlength{\aboverulesep}{2pt}
    \setlength{\tabcolsep}{5pt}
    \small
    \caption{\textbf{Summary of uncovered bugs. CVEs are anonymized for submission.}}
    \label{tab::vulnerabilities}
    \scriptsize{
        \begin{tabular}{c l l l}
            \toprule[1.5pt]
            \textbf{Target} & \textbf{Bug ID} & \textbf{Bug Type} & \textbf{Status} \\
            \toprule[1.5pt]
            \multirow{4}{*}{Zephyr \cite{Zephyr::Zephyr::Web}}  & \#1 & Out-of-bounds Write & Fixed \\
                                                                & \#2 & NULL Pointer Dereference & Fixed \\
                                                                & \#3 & Race Condition & Acknowledged \\
                                                                & \#4 & NULL Pointer Dereference & Unanswered \\
            \midrule
            \multirow{3}{*}{NuttX \cite{NuttX::NuttX::Web}} & \#5 & Out-of-bounds Write & \multirow{3}{*}{Fixed, CVE-2025-3xxxx} \\
                                                            & \#6 & Out-of-bounds Write & \\
                                                            & \#7 & Type Confusion & \\
            \midrule
            \multirow{3}{*}{\tabincell{c}{Mynewt/\\NimBLE \cite{Mynewt::Mynewt::Web}}} & \#8 & Out-of-bounds Write & Fixed, CVE-2025-5xxx0\\
                                            & \#9 & NULL Pointer Dereference & \multirow{2}{*}{Fixed, CVE-2025-5xxx7}\\
                                            & \#10 & NULL Pointer Dereference & \\
            \bottomrule[1.5pt]
        \end{tabular}
    }
\end{table}

\subsection{Security Implications -- Fuzzing RTOSes}
To answer \textbf{RQ 4}, we use the generated emulators to fuzz the Bluetooth host stacks of three popular RTOSes (Zephyr, NuttX and Mynewt/NimBLE) to demonstrate \sys's implications in firmware security. Compared to existing register-based firmware fuzzers \cite{Scharnowski::FuzzWare::Sec22, scharnowski_hoedur_2023}, \sys's accurate peripheral emulation helps the fuzzer to avoid spurious execution states and find more bugs. Existing firmware fuzzers treat peripheral hardware as a black box and use random values to emulate hardware behaviors. For example, they substitute register read results with random bytes, and trigger interrupts based on random selectors. However, due to constraints posed by hardware designs, certain register values may never appear, and interrupts cannot be triggered at any time. Since firmware typically relies on hardware behaviors, the inaccurate peripheral emulation described above can lead the firmware into spurious states, ultimately causing false crashes.




We implement the fuzzer based on LibAFL \cite{Fioraldi::LibAFL::CCS22}, with $\sim9,000$ lines of Rust and C code. For each target, we compile a firmware image based on official code samples, and feed test cases through the emulated hardware interfaces (i.e., data registers) during fuzzing. To prevent that the triggered crashes block the execution of remaining code, we adopt an iterative fuzzing strategy for our evaluation. For each iteration, we fuzz the target for 12 hours, inspect the triggered crashes, locate the corresponding bugs and fix them, and start the next iteration. We abort the iteration when no crashes are triggered during fuzzing. In our experiments, the fuzzing process aborts after three iterations for all targets.

In total, we uncovered 10 previously unknown bugs as shown in \autoref{tab::vulnerabilities}, all of which have been reported to the corresponding vendors. As of the time of writing, 8 bugs have been fixed (6 of which are assigned with CVEs), one bug has been acknowledged, while the remaining one bug has not yet been responded to. The 10 bugs are located in either peripheral drivers (\#1, \#5, and \#8) or protocol stacks. Since protocol stacks receive inputs from peripheral data registers, emulated peripherals are essential for uncovering these bugs. Some of these bugs (e.g., \#5 and \#6) allow attackers to control a large amount of memory and potentially have high security impacts.

The accurate emulation brought by \sys allows the fuzzer to uncover bugs that prior firmware fuzzers cannot find. These fuzzers do not fully respect the behaviors of actual hardware, and may cause spurious firmware execution states and false crashes. To validate this, we use Fuzzware \cite{Scharnowski::FuzzWare::Sec22} and Hoedur \cite{scharnowski_hoedur_2023}, two state-of-the-art firmware fuzzers to fuzz the three targets and inspect whether they can find the bugs listed in \autoref{tab::vulnerabilities}. We run each fuzzing campaign for 24 hours and repeat each campaign 5 times. As a result, these fuzzers often trigger false crashes and find none of the bugs.

We conduct case studies to investigate the root causes of the above phenomenon and confirm that inaccurate peripheral emulation has hindered prior firmware fuzzers. In both Zephyr and NuttX, the firmware uses not only the Nested Vectored Interrupt Controller (NVIC) but also peripheral-specific registers to enable and disable peripheral interrupts. For example, the STM32F4 UART RX interrupt can be enabled/disabled by writing 1/0 into the \texttt{RXNEIE} field of \texttt{CR} register. However, prior fuzzers fire interrupts by inspecting only the NVIC and randomly triggering those marked as enabled \cite{Scharnowski::FuzzWare::Sec22, scharnowski_hoedur_2023}. As a result, if an interrupt is enabled via the NVIC but subsequently disabled through a peripheral-specific register, these fuzzers may still trigger it. Notably, both Zephyr and NuttX use peripheral-specific registers to manage interrupts in critical sections to prevent race conditions. When a fuzzer triggers a disabled interrupt, it corrupts the protected critical section and causes false crashes. While for Mynewt/NimBLE, prior fuzzers cause false crashes because they do not properly emulate the number of interrupt priority bits in NVIC. As a result, they misinterpret the interrupt priorities and cause incorrect interrupt scheduling orders, which breaks Mynewt's task scheduler. We provide a more detailed case study in \refappendix{appendix::fuzz}. These case studies highlight the need for accurate peripheral emulation in the context of firmware fuzzing.



\section{Discussion}
\label{sec::discussion}

In this section, we further discuss the possibility to extend \sys to support more peripheral description materials and emulator frameworks , the involved manual efforts, and possible future directions in automated peripheral emulation.

\noindent\textbf{New Frontend and Backend.} Since \sys's design is centered around abstract models, it's possible to add more frontend and backend to support more peripheral description materials and more emulation frameworks. For example, one can implement a new frontend to extract model arguments from hardware manuals instead of driver source code, as well as a new backend to generate emulator code for other popular emulation frameworks such as Unicorn \cite{Quynh::Unicorn::BHUSA15} and Avatar \cite{Zaddach:Avatar::NDSS14}. Implementing a new backend requires solely engineering efforts. However, implementing a hardware manual oriented frontend for \sys can be challenging. Unlike driver source code, hardware manuals are unstructured and typically contain thousands of pages. Given that current LLMs only have a limited number of input tokens (e.g., 1,000,000 for Gemini 1.5 Flash), feeding the entire documents to the LLM is impractical. Besides, the code analysis-based resolution scheme proposed in this paper cannot be utilized to resolve ill-formed but benign responses, making it hard for the machine to understand LLM responses.
Due to these constraints, we leave the design and implementation of a new frontend as future work. 

\noindent\textbf{Involved Manual Efforts.} We try our best to make \sys as automatic as possible. For example, the resolution and validation schemes introduced in \S\ref{subsec::design::frontend} greatly help the automatic interaction between the machine and the LLM. However, although being minimal, it still requires some manual effort to use \sys to generate MCU emulators. First, the user must manually find corresponding driver source code, compile it to generate a compilation database, and select driver files as LLM inputs. MCU drivers can be easily obtained because MCU vendors typically provide open source drivers \cite{Lei::Perry::Sec24}, and compilation databases can be generated using existing tools (see \S\ref{subsec::experiment-setup}). Driver source files and their included header files contain sufficient context for the LLM's reasoning process. To select driver files, we first identify source files for a peripheral, then obtain header files included by them using the compiler's preprocessor, and filter out standard library headers and irrelevant headers (e.g,. header files for other peripherals). The remained headers, as well as source files, will serve as LLM inputs. This process is also lightweight and can be manually achieved within minutes. Second, due to the limitations of LLMs, the generated emulators might contain wrong hardware functionalities or miss certain functionalities, which requires manual fixes. However, as shown in \S\ref{subsec::consistency} and \S\ref{subsec::universality}, \sys can effectively filter out invalid responses and generate functionally correct emulators with a very high probability (e.g., 98.48\% unit test passing rate). Even when there are errors, fixing them typically requires minimal manual effort (e.g., 1 LoC). Finally, the required engineering efforts to support new peripheral models is minimal as shown in \S\ref{subsec::extensibility}, and is a one-time effort - the model can be used to generate emulator code for various peripheral implementations. Considering that developing an emulator for a single MCU typically requires thousands of lines of code, we believe the required manual efforts are acceptable and worthwhile.

\noindent\textbf{LLMs for Peripheral Emulation.} The emergence of LLMs has made it possible to solve problems that were difficult to address using traditional methods, particularly those involving complex semantic understanding. Peripherals contain complex semantics and functionalities. \sys breaks the overall peripheral into primitives, describes these primitives, and asks the LLM to determine primitive instances. A possible future direction could be using LLMs to decompose peripherals into various primitives, thus further automating the whole peripheral emulation process.

\noindent\textbf{Differences with Perry.} Although both \sys and Perry \cite{Lei::Perry::Sec24} generate emulator code, \sys has a completely different design and requires fewer inputs. Perry relies on ARM CMSIS System View Description (SVD) files to obtain structural (e.g., registers, fields) information and extracts four kinds of peripheral behaviors from driver source code to infer pre-defined semantic information for peripheral emulation. In contrast, \sys extracts all required information solely from driver source code. This fundamental difference distinguishes \sys from Perry in the following aspects.
First, \sys can cover more peripherals because it provides general peripheral representations to model various peripherals, while Perry only captures four kinds of peripheral behaviors that barely cover character peripherals like UART. Second, \sys has better extensibility due to the integration of LLMs. The two-fold modeling mechanism allows the users of \sys to describe peripheral semantics with natural language, and the LLM can directly extract required information based on these descriptions. In contrast, Perry relies on heuristic rules to drive the static analysis to capture peripheral behaviors, and its users have to apply manual annotations or even develop new static analysis rules to capture more peripheral behaviors. Third, \sys can handle hardware behaviors that Perry cannot handle. Perry relies on program analysis techniques and cannot capture hardware behaviors when they don't incur obvious control-flow or data-flow patterns (which Perry refers to as \textit{``implicit assumptions on hardware''}). However, \sys can capture such behaviors due to its LLM-assisted semantic understanding ability. \looseness=-1

\section{Related Work}
\label{sec::relatedwork}

\noindent\textbf{Static Firmware Analysis.} Due to the challenge of dynamically executing firmware, various static firmware analysis techniques have been proposed to find firmware bugs and vulnerabilities \cite{Davidson::FIE::Sec13, Shoshitaishvili::Firmalice::NDSS15, Hernandez::FirmUSB::CCS17, Redini::Karonte::SP20}. Such methods do not depend on firmware execution environments, thus eliminating the requirement for firmware rehosting techniques. However, they also inherit the drawbacks of static analysis techniques (e.g., false positives).

\noindent\textbf{Hardware-based Firmware Rehosting and Analysis.} Dynamic analysis techniques have shown great vulnerability discovery ability. To take advantage of these techniques, firmware rehosting techniques are proposed to create virtual firmware execution environments on host machines to reuse existing analysis tools. Hardware-based approaches assume the presence of actual physical devices and can be divided into two categories: hardware-in-the-loop and record-and-replay. Hardware-in-the-loop approaches \cite{Zaddach:Avatar::NDSS14, Koscher::SURROGATES::WOOT15, Corteggiani::Inception::Sec18} emulate CPU instructions on the host machine to execute firmware and forward peripheral accesses to actual devices. The second kind of approach \cite{Gustafson::Pretender::RAID19, Spensky::Conware::AsiaCCS21} instruments the firmware, executes it on a physical device, and records its peripheral access traces. These traces are then analyzed and replayed during firmware emulation. However, these techniques are either slow or have low fidelity, restricting analysis performance. To solve this problem, recent approaches \cite{Li::uAFL:ICSE22, Mera::SHiFT::SEC24, Liu::CO3::SEC24, Shi::IPEA::NDSS24} use the original hardware platform to perform various analyses and offload computing tasks to connected workstations to alleviate the computational and memory pressure on the MCU. However, as the size of the analyzed firmware grows, the additionally introduced resource consumption by existing dynamic analysis techniques (e.g., ASAN \cite{Serebryany::ASAN::ATC12}) becomes overwhelming for resource-constrained MCUs, making these methods infeasible.

\begin{sloppypar}
\noindent\textbf{Emulation-based Firmware Rehosting and Analysis.} To address the above limitations, emulation-based techniques have been proposed to create a fully virtualized execution environment \cite{zhou_iot_2025}. Firmware-dependent emulation techniques provide different virtual execution environments for different firmware and can be performed at either register level or function level. Function-level approaches \cite{Chen::Firmadyne::NDSS16, kim:2020:firmae, Jiang::ECMO::CCS21, Clements::HALucinator::Sec20, Maier::BaseSAFE::WiSec20, Li::Para-rehosting::NDSS21, Seidel::SAFIREFUZZ::SEC23, Hofhammer::SURGEON::BAR24} replace functions performing hardware interactions with customized ones to avoid hardware interactions while retaining the original semantics. Register-level methods \cite{Cao::Laelaps::ACSAC20, Zhou::uEmu::Sec21, Johnson::Jetset::Sec21, Liu::FirmGuide::ASE21, Feng::P2IM::Sec20, Mera::DICE::SP20, Scharnowski::FuzzWare::Sec22, scharnowski_hoedur_2023, chesser_multifuzz_2024} infer proper values for register accesses based on access context to bypass road blockers and explore firmware execution paths. However, since these methods cannot faithfully replicate hardware functionalities, they suffer from low universality and low fidelity, leading to infeasible execution paths or false positive crashes. Function-level and register-level approaches can be combined to emulate complex firmware such as TrustZone OS \cite{Harrison::PartEmu::Sec20}, Bluetooth firmware \cite{Ruge::Frankenstein::Sec20}, and baseband firmware \cite{Hernandez::FirmWire::NDSS22}. Firmware-agnostic emulation techniques manage to solve these problems by faithfully emulating MCU hardware. To automate this process, existing approaches summarize a limited set of heuristic rules to approximate peripheral behaviors and extract these rules from hardware manuals \cite{Zhou::sEmu::CCS22} or driver source code \cite{Lei::Perry::Sec24}. Although recent efforts \cite{Lei::Perry::Sec24} have made promising advances in this direction, the rule-based mechanism has made such methods less practical due to the variety of peripherals and their functionalities.
\end{sloppypar}

\section{Conclusion}
\label{sec::conclusion}

In this paper, we propose \sys, a flexible MCU peripheral emulation framework. Through a two-fold peripheral modeling mechanism, \sys builds abstract models for various categories of MCU peripherals, with peripheral-specific implementation details left as model parameters. \sys adopts a LLM-based frontend to extract model arguments for different peripheral implementations from driver source code, instantiates model instances, and uses a backend to generate corresponding emulation code using template-based code synthesis. \sys is highly extensible and now supports 12 kinds of MCU peripherals. Evaluation results on 90 firmware samples across 15 MCUs demonstrate the fidelity and universality of the generated MCU emulators with a 98.48\% unit test passing rate, surpassing state-of-the-art. We also use the generated emulators to fuzz the Bluetooth host stacks of three popular RTOSes and find 10 previously unknown bugs, demonstrating \sys's implications on firmware security.

\paragraph{Availability} \sys is available at \url{https://github.com/FlexEmu/flexemu}.
\bibliographystyle{ACM-Reference-Format}
\bibliography{reference}

\appendix

\newcounter{promptcounter}
\newtcolorbox[use counter=promptcounter,number format=\arabic]{promptbox}[2][]{
    enhanced jigsaw,
    colback=yellow!6,
    colbacktitle=yellow!95!black,
    breakable,
    fontupper=\footnotesize\sffamily\itshape,
    title={\textcolor{black}{\small{\textbf{Prompt~\thetcbcounter: #2}}}},
    #1
}
\newcommand{\promptcounterautorefname}{Prompt}

\newtcolorbox[]{sysinstbox}[2][]{
    enhanced jigsaw,
    colback=yellow!6,
    colbacktitle=yellow!95!black,
    breakable,
    fontupper=\footnotesize\sffamily\itshape,
    title={\textcolor{black}{\small{\textbf{#2}}}},
    #1
}

\section{The System Instruction and Prompt Templates Used by \sys}
\label{appendix::prompt-template}

We list the used system instruction and prompt templates in this section. Note that named placeholders enclosed in \{\} are replaced by real contents at runtime.

\begin{sysinstbox}{System Instruction.}
You are an expert driver code analyzer, and your job is to answer the user's query based on the driver code files you have access to.
\end{sysinstbox}

\begin{promptbox}[label=prompt::pcat]{Peripheral Category Identification (Stage \circlenum{1}).}
There are 12 abstract peripheral categories: [ADC, DAC, DMA, Ethernet, GPIO, RNG, SDHC, SDIO, SPI, I2C, Timer, UART]. Find all peripheral categories for the \{MCU\_NAME\} MCU and output in JSON format:

[{"<peripheral category name>": "<abstract category>"},...]
\end{promptbox}

\begin{promptbox}[label=prompt::reg]{Register Identification (Stage \circlenum{2}).}
Find all registers of the \{PERIPHERAL\_NAME\} peripheral. Output in JSON format like this:

\{``regs'': [\{

\ \ ``name'': ``<register name>'',

\ \ ``width'': ``<register width in bits>'',

\ \ ``offset'': ``<address offset within the peripheral>''\}, ...]\}

Think step by step.
\end{promptbox}

\begin{promptbox}[label=prompt::field]{Register Field Identification (Stage \circlenum{3}).}
Find all fields of the \{REGISTER\_NAME\} register of the \{PERIPHERAL\_NAME\} peripheral. Output in JSON format like this:

\{``fields'': [\{

\ \ ``name'': ``<field name>'',

\ \ ``pos'': ``<bit position of the field within the register>'',

\ \ ``width'': ``<field width in bits>''\}, ...]\}


Think step by step.
\end{promptbox}

\begin{promptbox}[label=prompt::update]{Update Identification (Stage \circlenum{4}).}
When the driver sets/clears some register fields (condition), hardware may take actions and sets/clears some register fields (action). To wait the hardware to finish, the driver polls for these register fields. The above procedure looks like this:

SET(REG\_A, FIELD\_A) or CLEAR(REG\_A, FIELD\_A) // condition

SET(REG\_B, FIELD\_B) or CLEAR(REG\_B, FIELD\_B) // condition, can bemultiple

while ((REG\_C \& FIELD\_C) == 0/1); // action

while ((REG\_D \& FIELD\_D) == 0/1); // action, can be multiple

For the \{PERIPHERAL\_NAME\} peripheral, find all such situations. Output in JSON format like this:

\{``updates'': [\{``condition'': [...], ``action'': [...]\}]\}


Think step by step.
\end{promptbox}

\begin{promptbox}[label=prompt::semantic]{Peripheral Semantic Identification (Stage \circlenum{5}).}
Summarize information about the \{PERIPHERAL\_NAME\} peripheral and output in JSON format like this:

\{JSON\_LIKE\_PROMPT\}

Think step by step.
\end{promptbox}

\begin{promptbox}[label=prompt::instance]{Peripheral Instance Identification (Stage \circlenum{6}).}
Find all peripheral instances of kind \{PERIPHERAL\_NAME\}. Output in JSON format like this:

``instances'': [\{

    "name": "<name of the instance>",

    "base": "<base address of the peripheral instance>",

    "irqs": ["<interrupt number>", ...]

  \},...]

Think step by step.
\end{promptbox}

\begin{promptbox}[label=prompt::irq]{Interrupt Association (Stage \circlenum{7}).}
Associate interrupt events listed in the given JSON with their interrupt numbers by filling the blanks.

\{\{

``instance'': ``<instance name>'',

``events'': [\{``event'': ``<event name>'', ``irq'': ``<BLANK: interrupt number>''\}, ...]

\}, ...\}

Think step by step.
\end{promptbox}


\begin{listing}[!]
\begin{minted}[
breakanywhere,
baselinestretch=1.0,
numbersep=2pt,
bgcolor=bgcolor,
fontsize=\scriptsize,
linenos,
breaklines,
highlightcolor=highlightcolor,
]{python}
class DMATransDescModel(BaseModel):
  enable: Swt = Field(description="when to enable the channel")
  complete: Evt = Field(description="the event generated when the transfer completes")
  src: Reg = Field(description="the register holding DMA transfer source address")
  src_width: RegFieldMap = Field(description="the register field representing source transfer chunk width (in bytes)")
  dst: Reg = Field(description="the register holding DMA transfer destination address")
  dst_width: RegFieldMap = Field(description="the register field representing destination transfer chunk width (in bytes)")
  cnt: Reg = Field(description="the register holding the number of data to be transferred")
  direction: Optional[RegFieldState] = Field(description="the register field representing transfer direction")

class DMAModel(PeripheralModel):
  trans_descs: list[DMATransDescModel] = Field(description="a list of transfer descriptors")
\end{minted}
\small
\caption{\textbf{DMA controller model definition.}}
\label{lst::dma-model}
\end{listing}

\section{Case Study -- Emulating DMA Controllers}
\label{appendix::dma-example}
We illustrate how to use \sys to model and emulate DMA controllers. DMA controllers manage data transfers between peripherals and the memory, using channels as the unit of operation. DMA controllers use hardware transfer descriptors consisted of registers to configure and control these channels. Typically, each transfer descriptor has a transfer source address, a transfer destination address, and the transfer data count. When the channel is enabled, the peripheral retrieves the source address and the destination address from corresponding registers. Then, it transfers the configured count of data from the source address to the destination address and generates an event upon completion. During this process, data are actually transferred in small chunks, and the transfer descriptor configures the chunk width (e.g., 4 bytes). Sometimes, the transfer direction can even be inverted to go from the destination address to the source address. Based on this semantic level abstraction, we build a semantic model for DMA controllers using \sys as shown in \autoref{lst::dma-model}. The DMA controller model consists of a list of transfer descriptors, each containing the above key elements that are represented by our primitives in \autoref{tab::primitive} and their semantic descriptions. Note that \texttt{Field()} is a function provided by the \texttt{pydantic} library and is used to attach functionality descriptions to  primitives.



Based on this model, we build an emulator template in the backend that mimics the hardware behaviors of DMA controllers. Specifically, the template starts transfer when the channel is enabled by writing corresponding registers, retrieves data from the source address based on the chunk width, and sends data to the destination address based on the chunk width. When the total transferred data count reaches the configured number, the template sets corresponding register fields and generates an event to indicate transfer completion.

\begin{listing}[!]
\begin{minted}[
breakanywhere,
baselinestretch=1.0,
numbersep=2pt,
bgcolor=bgcolor,
fontsize=\scriptsize,
breaklines,
highlightcolor=highlightcolor,
]{json}
{
  // a list of transfer descriptors
  "trans_descs": [
    {
      // when to enable the channel
      "enable": {
        "enable": {
          "reg": "<register name>", "field": "<field name>", "value": "<value of the field>",
        },
        "disable": {
          "reg": "<register name>", "field": "<field name>", "value": "<value of the field>",
        },
        "active": {
          "reg": "<register name>", "field": "<field name>", "value": "<value of the field>",
        },
      },
      // the event generated when the transfer completes
      "complete": {
        "happen": {
          "reg": "<register name>", "field": "<field name>",
          "value": "<when the event happens, the field is set to this value>",
        },
        "active": {
          "reg": "<register name>", "field": "<field name>",
          "value": "<when the event is enabled, the field is set to this value>",
        },
        "enable": {
          "reg": "<register name>", "field": "<field name>",
          "value": "<the event interrupt is enabled when this value is written into the field>",
        },
        "disable": {
          "reg": "<register name>", "field": "<field name>",
          "value": "<the event interrupt is disabled when this value is written into the field>",
        },
        "clear": {
          "reg": "<register name>", "field": "<field name>",
          "value": "<the event happen flag is cleared when this value is written into the field>",
        },
      },
      "src": "<the register holding DMA transfer source address>",
      // the register field representing source transfer chunk width (in bytes)
      "src_width": {
        "reg": "<register name>", "field": "<field name>",
        "map": {"<field value>": "<mapped value>", ...},
      },
      "dst": "<the register holding DMA transfer destination address>",
      // the register field representing destination transfer chunk width (in bytes)
      "dst_width": {
        "reg": "<register name>", "field": "<field name>",
        "map": {"<field value>": "<mapped value>", ...},
      },
      "cnt": "<the register holding the number of data to be transferred>",
      // [OPTIONAL] the register field representing transfer direction
      "dir": {
        "reg": "<register name>", "field": "<field name>", "value": "<field value>",
      },
    },
    ...,
  ],
},
\end{minted}
\small
\caption{\textbf{JSON-like template generated using the DMA model.}}
\label{lst::dma-template}
\end{listing}

\sloppypar
We then demonstrate a concrete example on how \sys generates emulators for STM32F1xx's DMA controller peripheral. \sys first applies \autoref{prompt::pcat} to driver source code and the LLM response \texttt{["DMA": "DMA", ...]} indicates that the MCU has a kind of peripheral named ``DMA'', and that the associated peripheral model is DMA. Then, \sys replaces \texttt{PERIPHERAL\_NAME} with \textit{``DMA''} in \autoref{prompt::reg} to ask the LLM to identify all registers of a DMA peripheral. For example, if the LLM returns \texttt{\{``regs'': [\{``name'': ``ISR'', ``width'': ``32'', ``offset'': ``0x00''\}, ...]\}}, then the peripheral has a register named \texttt{ISR} that is 32-bit long and is mapped at the address \texttt{0x00}. We create a \texttt{Reg} accordingly and add it to the peripheral model. \sys then iterates all identified registers, replaces \texttt{REGISTER\_NAME} with the register name in \autoref{prompt::field}, and asks the LLM to identify fields for each register. For example, if the LLM returns \texttt{\{``fields'': [\{``name'': ``GIF1'', ``pos'': ``0'', ``width'': ``1''\}, ...]\}} for the \texttt{ISR} register, then the register has a 1-bit long field named \texttt{GIF1} starting from bit 0. We create a \texttt{RegField} accordingly and set its \texttt{reg} member to the \texttt{ISR} register. After this, \sys then applies \autoref{prompt::update} to collect all \texttt{Update} of DMA peripherals. Since the driver code does not contain such patterns, the LLM returns an empty array: \texttt{\{``updates'': []\}}. Then, based on the model in \autoref{lst::dma-model}, \sys automatically generates the JSON-like prompt as shown in \autoref{lst::dma-template}, embeds it into \autoref{prompt::semantic}, and sends the complete prompt to the LLM. The LLM then returns the extracted information as shown in \autoref{lst::dma-llm-res}. We only show the extracted information for the first hardware transfer descriptor due to space constraints. \sys parses the response and creates a model instance accordingly. \sys then uses \autoref{prompt::instance} to identify all DMA peripherals in the MCU.  For example, if the LLM returns \texttt{\{``instances'': [\{``name'': ``DMA1'', ``base'': ``0x40020000'', ``irqs'': [``DMA1\_Channel1\_IRQn'', ...]\}]\}}, then the MCU only has one DMA peripheral called \texttt{DMA1} that is mapped at the address \texttt{0x40020000}. \texttt{DMA1} also has an interrupt number \texttt{DMA1\_Channel1\_IRQn}. Since \texttt{DMA1\_Channel1\_IRQn} is an expression, \sys uses the code analysis-based resolution technique described in \S\ref{subsec::design::frontend} to resolve it and gets the number \texttt{11}. Using such information, \sys obtains a peripheral instance by combining the model instance and the collected information. Finally, \sys asks the LLM to identify an interrupt number for each \texttt{Evt} contained in the model instance. To achieve this, \sys traverses the model instance and collects contained \texttt{Evt} members. \sys automatically assigns a name for each \texttt{Evt} based on the declared member name. For example, \texttt{complete} in the first transfer descriptor is named to \texttt{trans\_desc.0.complete}. \sys then assembles \autoref{prompt::irq} and sends it to the LLM. For example, the prompt looks like: \texttt{\{\{``instance'': ``DMA1'', ``events'': [\{``event'': ``trans\_desc.0.complete'', ``irq'': ``<BLANK: interrupt number>''\}, ...]\}\}}. The LLM's response may look like: \texttt{\{\{``instance'': ``DMA1'', ``events'': [\{``event'': ``trans\_desc.0.complete'', ``irq'': ``11''\}, ...]\}\}}, which means that when the transmission mission in the first descriptor finishes, the raised event will activate the 11th interrupt. Therefore, we set the \texttt{irq\_line} member of the corresponding \texttt{Evt} object to 11. After the above steps, \sys has obtained a complete \texttt{DMAModel} object and the associated instance objects. \sys's backend then generates emulators for these instances based on the emulator template.

\begin{listing}[!]
\begin{minted}[
breakanywhere,
baselinestretch=1.0,
numbersep=2pt,
bgcolor=bgcolor,
fontsize=\scriptsize,
breaklines,
highlightcolor=highlightcolor,
]{json}
{
  "trans_descs": [
    {
      "enable": {
        "enable": {"reg": "Channel_0_CCR", "field": "EN", "value": "0x01"},
        "disable": {"reg": "Channel_0_CCR", "field": "EN", "value": "0x00"}
      },
      "complete": {
        "happen": {"reg": "ISR", "field": "TCIF1", "value": "0x01"},
        "active": {"reg": "Channel_0_CCR", "field": "TCIE", "value": "0x01"},
        "enable": {"reg": "Channel_0_CCR", "field": "TCIE", "value": "0x01"},
        "disable": {"reg": "Channel_0_CCR", "field": "TCIE", "value": "0x00"},
        "clear": {"reg": "IFCR", "field": "CTCIF1", "value": "0x01"}
      },
      "src": "Channel_0_CMAR",
      "src_width": {
        "reg": "Channel_0_CCR", "field": "MSIZE",
        "map": {"0x00": "1", "0x01": "2", "0x02": "4"}
      },
      "dst": "Channel_0_CPAR",
      "dst_width": {
        "reg": "Channel_0_CCR", "field": "PSIZE",
        "map": {"0x00": "1", "0x01": "2", "0x02": "4"}
      },
      "cnt": "Channel_0_CNDTR",
      "dir": {"reg": "Channel_0_CCR", "field": "DIR", "value": "0x01"}
    },
    ...,
  ],
},
\end{minted}
\small
\caption{\textbf{Part of the LLM's response for the prompt generated from the DMA model.}}
\label{lst::dma-llm-res}
\end{listing}

\section{Case Study -- Emulating Ethernet Controllers}
\label{appendix::eth}
We list the peripheral model of Ethernet controllers in \autoref{lst::eth-model}. \sys takes similar steps as shown in \refappendix{appendix::dma-example} to generate emulators for Ethernet controllers.

\begin{listing}[!]
\begin{minted}[
breakanywhere,
baselinestretch=1.0,
numbersep=2pt,
bgcolor=bgcolor,
fontsize=\scriptsize,
linenos,
breaklines,
highlightcolor=highlightcolor,
]{python}
class EthTransDescAddressingMethod(Enum):
  LinkedList = auto()
  Array = auto()

class EthTransDescModel(BaseModel):
  trans_desc_struct: str = Field(description="name of the transfer descriptor struct")
  tx_frame_len: MemField = Field(description="the field within the transfer descriptor struct that holds the number of bytes to be transmitted in a frame")
  rx_frame_len: MemField = Field(description="the field within the transfer descriptor struct that holds the number of received bytes in a frame")
  buf: MemField = Field(description="the field within the transfer descriptor struct that holds the buffer address"
  addr_method: EthTransDescAddressingMethod = Field(description="descriptor addressing method")
  last_rx_seg: MemFieldState = Field(description="when the field is set to this value, the corresponding descriptor represents the last received segment")
  last_tx_seg: MemFieldState = Field(description="when the field is set to this value, the corresponding descriptor represents the last segment to be transmitted")
  own: MemFieldState = Field(description="when the field is set to this value, the corresponding descriptor can be manipulated by the hardware")
  first_rx_seg: Optional[MemFieldState] = Field(description="when the field is set to this value, the corresponding descriptor represents the first received segment")
  rx_buf_len: Optional[MemField] = Field(description="the field within the transfer descriptor struct that holds the length of `buf`")
  next: Optional[MemField] = Field(description="the field within the transfer descriptor struct that holds the address of the next descriptor. only present when addressing method is `LinkedList`")
  last_desc: Optional[MemFieldState] = Field(description="when this is the last descriptor in the array, the corresponding descriptor represents the first received segment. only present when addressing method is `Array`")

class EthModel(PeripheralModel):
  trans_desc: EthTransDescModel = Field(description="a list of transfer descriptors")
  rx_desc_reg: Reg = Field(description="the register holding the address of rx descriptors")
  tx_desc_reg: Reg = Field(description="the register holding the address of tx descriptors")
  rx_enable: Swt = Field(description="when to enable rx")
  tx_enable: Swt = Field(description="when to enable tx")
  rx_done: Evt = Field(description="the event generated when a frame is received")
  tx_done: Evt = Field(description="the event generated when a frame is transmitted")
  rx_buf_len_reg: Optional[Reg] = Field(description="the register holding the length of `trans_desc.buf`, can be stored here or in `trans_desc.rx_buf_len`")
\end{minted}
\small
\caption{\textbf{Ethernet controller model definition.}}
\label{lst::eth-model}
\end{listing}

\section{Case Study -- How Inaccurate Peripheral Emulation Blocks Firmware Fuzzing}
\label{appendix::fuzz}

We take Zephyr as an example to illustrate how inaccurate peripheral emulation blocks existing firmware fuzzers. Zephyr's Bluetooth stack takes inputs and sends messages using the H4 driver as listed in \autoref{lst::zephyr-h4-driver}. Whenever there is incoming data, the driver processes it via the UART interrupt handler \texttt{bt\_uart\_isr}. If the accumulated data has formed a complete packet, the handler puts the packet into a queue (i.e., \texttt{h4->rx.fifo}). To process these packets, the driver executes \texttt{rx\_thread} within a dedicated thread. \texttt{rx\_thread} retrieves the packet from the queue and finally sends it to the stack (line 18-21). Since the driver uses a global context object (i.e., \texttt{h4}) to manage the above process, the driver actively enables (line 17) and disables (line 23) the UART interrupt to guard critical sections and prevent racing conditions. Note that the driver configures UART registers instead of the NVIC to enable/disable interrupts.

The emulators adopted by existing firmware fuzzers collect active interrupts based on NVIC. Therefore, when the firmware executes within these emulators, they do not know that the UART interrupt has been disabled after line 23, and the fuzzer can still fire the UART interrupt. However, this corrupts the critical section and can directly crash the firmware. For example, if the fuzzer fires the UART interrupt after line 14, line 5 can be executed to set \texttt{h4->rx.buf} to a null pointer. After the interrupt handler returns to resume the execution of \texttt{rx\_thread}, the following code may directly dereference the null pointer and cause the firmware to crash. In contrast, the emulators generated by \sys are aware of the interrupt enabling and disabling operations due to the properly emulated peripheral hardware.

\begin{listing}[!]
\begin{minted}[
breakanywhere,
baselinestretch=1.0,
numbersep=2pt,
bgcolor=bgcolor,
fontsize=\scriptsize,
breaklines,
linenos,
highlightcolor=highlightcolor,
highlightlines={11-17,24-25},
]{c}
// call chain: bt_uart_isr -> process_rx -> read_payload
static inline void read_payload(...) {
  ...
  struct net_buf *buf = h4->rx.buf;
  h4->rx.buf = NULL;
  ...
  k_fifo_put(&h4->rx.fifo, buf);
}

static void rx_thread(...) {
  ... // uart irq is disabled by default
  while (1) {
    if (h4->rx.have_hdr && !h4->rx.buf) {
      h4->rx.buf = get_rx(h4, K_FOREVER);
      ...
    }
    uart_irq_rx_enable(cfg->uart); // set CR1[RXNEIE]
    struct net_buf *buf = k_fifo_get(&h4->rx.fifo, K_FOREVER);
    do {
      uart_irq_rx_enable(cfg->uart);
      h4->recv(dev, buf);
      k_yield();
      uart_irq_rx_disable(cfg->uart); // clear CR1[RXNEIE]
      ...
    } while (...);
  }
}
\end{minted}
\small
\caption{\textbf{Bluetooth H4 driver of Zephyr. Highlighted lines are within the same critical section.}}
\label{lst::zephyr-h4-driver}
\end{listing}

\end{document}